%% file: paper.tex
\newif\iffinal
\newif\ifarxiv
\arxivtrue
\finaltrue

\documentclass[sigplan,10pt,nonacm]{acmart}

\usepackage{tikz}
\usetikzlibrary{arrows.meta}
\usepackage{subfig}
\usepackage{multirow}
\usepackage{xspace}
\usepackage{algorithm}
\usepackage[noend]{algpseudocode}
\usepackage{afterpage}
\usepackage{mathtools}
\usepackage{paralist}
\usepackage{booktabs}
\definecolor{knop_color}{HTML}{70AD47}
\definecolor{tbop_color}{HTML}{5B9BD5}
\definecolor{tbgraph_bg}{HTML}{DEEBF7}
\definecolor{kngraph_bg}{HTML}{E2EFDA}
\definecolor{pipebg}{HTML}{FFFFDF}
\definecolor{pipepass}{HTML}{C8E6C9}
\definecolor{pipefail}{HTML}{FFCDD2}
\definecolor{pipebox}{HTML}{E8EAF6}
\definecolor{pipearrow}{HTML}{546E7A}
\usepackage{makecell}
\usepackage{threeparttable}
\usepackage[nosort]{cleveref}

\newmuskip\normalthickmuskip
\newmuskip\normalthinmuskip
\AtBeginDocument{%
  \normalthickmuskip=\thickmuskip
  \normalthinmuskip=\thinmuskip
}
\everymath{%
  \thickmuskip=\muexpr\normalthickmuskip-(1\normalthinmuskip plus 1\normalthinmuskip)\relax
}
\everydisplay{\thickmuskip=\normalthickmuskip}
\let\texdisplaystyle\displaystyle
\renewcommand{\displaystyle}{\texdisplaystyle\the\everydisplay}

\algrenewcommand\algorithmicindent{0.5em}
\algrenewcommand\algorithmicrequire{\textbf{Input:}}
\algrenewcommand\algorithmicensure{\textbf{Output:}}

\newcommand{\ZJ}[1]{}
\newcommand{\MW}[1]{}
\newcommand{\oded}[1]{}

\ifarxiv
\newcommand{\Sys}{Prism\xspace}
\newcommand{\sys}{Prism\xspace}
\else
\newcommand{\Sys}{SIGMA\xspace}
\newcommand{\sys}{SIGMA\xspace}
\fi

\newcommand{\commentout}[1]{}
\algnewcommand{\LeftComment}[1]{\Statex \(\triangleright\) #1}
\newcommand{\er}[1]{\mbox{\rm\em #1}}

\newcommand{\removed}[1]{}
\renewcommand\footnotetextcopyrightpermission[1]{}

\iffinal

\else

\fi

\crefname{part}{\S}{\S\S}
\crefname{chapter}{\S}{\S\S}
\crefname{section}{\S}{\S\S}
\crefname{subsection}{\S}{\S\S}

\newcommand{\imap}{\er{imap}\xspace}
\newcommand{\omap}{\er{omap}\xspace}
\newcommand{\fmap}{\er{fmap}\xspace}
\newcommand{\graph}{sGraph\xspace}
\newcommand{\graphs}{sGraphs\xspace}

\title{\sys: Symbolic Superoptimization of Tensor Programs}

\author{Mengdi Wu}
\affiliation{\institution{Carnegie Mellon University}\country{USA}}
\email{mengdiwu@andrew.cmu.edu}

\author{Xiaoyu Jiang}
\affiliation{\institution{Tsinghua University}\country{China}}
\email{jiangxia22@mails.tsinghua.edu.cn}

\author{Oded Padon}
\affiliation{\institution{Weizmann Institute of Science}\country{Israel}}
\email{oded.padon@weizmann.ac.il}

\author{Zhihao Jia}
\affiliation{\institution{Carnegie Mellon University}\country{USA}}
\email{zhihao@cmu.edu}

\begin{document}
\settopmatter{printfolios=true}

\input{abstract}

\maketitle

\input{intro}

\input{mlcg}

\input{generate}
\input{verification}

\input{instantiation}

\input{impl}

\input{eval2}
\input{related}

\section{Conclusion}
\label{sec:conclusion}

We presented \sys, the first symbolic superoptimizer for tensor programs. The key idea behind \sys is \graphs, a symbolic graph representation that compactly encodes large families of tensor programs by abstracting mappings and parallelization parameters as symbolic variables. This enables \sys to decouple graph structure search from mapping enumeration and parameter tuning, significantly reducing the search space compared to concrete enumeration. We introduced symbolic dimension matching and expression-guided pruning to efficiently navigate this space, and developed an axiom-based verification framework to check functional equivalence of symbolic graphs. Our evaluation on workloads from modern LLMs shows that \sys outperforms existing systems by up to $2.2\times$ over state-of-the-art superoptimizers while reducing optimization time by up to $3.4\times$.

\begin{acks} 
This research is partially supported by NSF awards CNS-2211882 and CNS-2239351, a Sloan research fellowship, and research awards from Amazon, Cisco, Google, Jane Street, Meta, NVIDIA, Oracle, Qualcomm, and Samsung. Mengdi Wu is supported by the Amazon AI Fellowship.
This research is partially supported by Israel Science Foundation research grant (ISF's No.\ 4136/25) and the Maimonides Fund's Future Scientists Center,
a research grant from the Center
for New Scientists at the Weizmann Institute of Science, and a grant from the Azrieli Foundation.
\end{acks}

\bibliographystyle{ACM-Reference-Format}
\bibliography{bibliography}

\end{document}

%% file: abstract.tex
\begin{abstract}
This paper presents \sys, the first \emph{symbolic superoptimizer} for tensor programs. The key idea is \graph, a symbolic, hierarchical representation that compactly encodes large classes of tensor programs by symbolically representing some execution parameters. \sys organizes optimization as a two-level search: it constructs symbolic graphs that represent families of programs, and then instantiates them into concrete implementations. This formulation enables structured pruning of provably suboptimal regions of the search space using symbolic reasoning over operator semantics, algebraic identities, and hardware constraints. 

We develop techniques for efficient symbolic graph generation, equivalence verification via e-graph rewriting, and parameter instantiation through auto-tuning. Together, these components allow \sys to bridge the rigor of exhaustive search with the scalability required for modern ML workloads. Evaluation on five commonly used LLM workloads shows that \sys achieves up to $2.2\times$ speedup over best superoptimizers and $4.9\times$ over best compiler-based approaches, while reducing end-to-end optimization time by up to $3.4\times$.

\end{abstract}

%% file: intro.tex
\section{Introduction}
\label{sec:intro}
Efficient execution of ML models on GPUs is fundamental to modern AI applications. Today's ML systems generally express computations as {\em tensor programs}, typically represented as directed acyclic graphs (DAGs), where nodes correspond to tensor operators and edges denote tensors (i.e., multi-dimensional arrays).
Most existing ML systems optimize tensor programs through transformation rules and scheduling templates manually designed by domain experts. Systems such as TensorFlow, TensorRT, and TVM incorporate hand-crafted graph rewrite rules and operator fusion heuristics~\cite{Tensorflow, tensorrt, tvm}, while vendor-provided kernel libraries such as cuDNN and cuBLAS offer highly optimized implementations for a fixed set of operators~\cite{cudnn, cublas}.
Although effective for widely used operations and models, this paradigm has two fundamental limitations.
First, it requires substantial engineering effort to support new operators or hardware targets. 
For example, adapting optimized kernels such as FlashAttention~\cite{dao2023flash} to a new GPU architecture generally requires months of manual tuning and engineering.
Second, manually designed rules explore only a limited portion of the optimization space. Human intuition is inherently insufficient to capture the combinatorial interactions among algebraic transformations, data layouts, and hardware-specific scheduling decisions.

\emph{Superoptimization} has emerged as a promising paradigm for automatically discovering fast tensor programs without relying on manually specified rules. Originating from the compiler literature, superoptimization explores (exhaustively or heuristically) a search space of candidate programs and retains those that are both functionally equivalent to the original and empirically faster.
TASO~\cite{TASO} pioneered this approach for tensor programs by automatically generating graph substitutions and then applying them to optimize tensor programs: it enumerates small computation subgraphs over a predefined operator set, identifies equivalent pairs via a combination of random testing and formal verification, and applies the generated transformation to optimize the target program using a cost-guided search.
Mirage~\cite{mirage} applies superoptimization across multiple levels of the GPU execution hierarchy through $\mu$Graphs, a unified representation that captures optimizations at the kernel, thread-block, and thread levels. This multi-level search enables coordinated algebraic and scheduling transformations, including the synthesis of entirely new custom kernels beyond the reach of single-level approaches.

Recently, large language models (LLMs) have been used to generate optimized GPU kernels~\cite{wiedemann2026kernelfoundry, wei2025astra}.
More broadly, AlphaEvolve has been proposed as a general-purpose superoptimizer that leverages LLMs to guide the search~\cite{novikov2025alphaevolve}. It uses an evolutionary framework in which LLMs 
iteratively propose and refine code candidates that are validated by automated evaluators.
In a variety of well-defined tasks, AlphaEvolve demonstrates that LLM-guided search can discover optimizations surpassing both human-engineered and prior automated solutions. However, its applicability to optimizing tensor programs has not been explored yet.

While {\em enumeration}-based superoptimizers like TASO and Mirage achieve strong performance, their reliance on exhaustive search introduces a fundamental scalability bottleneck: the number of candidate programs grows combinatorially with the number of operators and levels in the execution hierarchy, making exhaustive enumeration impractical for large or deeply nested programs. 
On the other hand, {\em sampling}-based superoptimizers such as AlphaEvolve use learned priors to guide the search, enabling exploration over substantially larger spaces. However, these methods treat the optimization landscape as largely unstructured, which can lead to unstable search behavior and provides limited guarantees on coverage or completeness of the explored program space.

We introduce \sys, the first \emph{symbolic superoptimizer} for tensor programs. Rather than enumerating concrete candidate programs via brute-force enumeration or stochastic sampling, \sys organizes the search space into a two-level hierarchy. At the upper level, it constructs a \emph{symbolic} graph representation, called \graph, which compactly encodes entire families of tensor programs; at the lower level, each \graph is instantiated into many concrete implementations.

This symbolic formulation enables \sys to explore substantially larger search spaces and discover higher-quality optimizations compared to prior enumeration- and sampling-based approaches for two key reasons. First, \graph encodes operator semantics, algebraic identities, and hardware constraints as symbolic expressions, allowing \sys to prune provably suboptimal regions of the search space before materializing concrete programs. This structured pruning enables scalability to optimization problems that are intractable for exhaustive enumeration.
Second, \sys preserves optimality guarantees: the pruning process is sound and does not eliminate optimal solutions. This property fundamentally distinguishes symbolic superoptimization from prior sampling-based approaches.
Overall, \sys bridges the rigor of exhaustive search and formal verification with the scalability requirements of modern ML workloads.

\paragraph{Symbolic graph representation.}
A key idea of \sys is \graphs, a \emph{symbolic, hierarchical} graph representation of tensor programs. Unlike conventional representations that fix execution parameters (e.g., grid dimensions, block dimensions, and tensor-to-thread mappings) to concrete values, \sys encodes these attributes symbolically while instantiating only the high-level computational structure. This design allows a single \graph to represent a large class of related programs, enabling  symbolic reasoning over the optimization space and more efficient exploration.

\paragraph{\graph generation.}
Given an input tensor program, \sys enumerates candidate \graphs whose instantiations may be functionally equivalent to the target program. A key challenge is to prune the search space both \emph{effectively} (i.e., eliminating a large fraction of invalid candidates) and \emph{efficiently} (i.e., with minimal overhead). To this end, we introduce two complementary pruning techniques. \emph{Symbolic dimension matching} ensures compatibility of operator dimensions when expressed as symbolic expressions, while  \emph{symbolic expression pruning} extends the abstract expression checking from previous work~\cite{mirage} to the symbolic setting. Together, these techniques significantly reduce the search space and enable \sys to scale to workloads where concrete enumeration becomes intractable (\S\ref{sec:search}).

\paragraph{\graph verification.}
For each candidate \graph, \sys must verify functional equivalence with the input program without committing to concrete parallelization parameters. Prior work relies on random testing, which requires fixed tensor shapes and is therefore incompatible with symbolic representations. Instead, \sys encodes both the input program and the candidate \graph as expressions over a set of predefined operators and performs equivalence checking using e-graphs~\cite{egg} under a set of algebraic axioms. These axioms capture the mathematical properties of supported operators and their interaction with parallelization, enabling verification of symbolic graphs independent of concrete parameter values (\S\ref{sec:optimizer}).

\paragraph{\graph instantiation.}
For each verified \graph, \sys instantiates the remaining symbolic parameters to produce optimized GPU kernels for a given target configuration. We employ random sampling with GPU profiling to tune these parameters, leveraging the effectiveness of auto-tuning techniques that have been extensively studied for tensor programs (\S\ref{sec:instantiation}).

\paragraph{Evaluation.}
We evaluate \sys on five workloads commonly used in modern LLM architectures, including fused normalization-linear layers, gated MLPs, and group-query attention. Across these workloads, \sys outperforms existing systems by up to $2.2\times$ over the state-of-the-art superoptimizers and $4.9\times$ over traditional compiler-based approaches, by discovering optimizations that require exploring a substantially larger space of parallelization strategies than prior approaches can handle. Meanwhile, \sys reduces end-to-end optimization time by up to $3.4\times$.

%% file: mlcg.tex
\section{Symbolic Graph Representation}
\label{sec:hcg}

\Sys uses \graph to compactly encode large classes of tensor programs by symbolically representing selected execution parameters. 
This section first reviews the GPU programming model, then formalizes symbolic parallelization parameters, mappings, and tensor shapes, and finally expresses correctness constraints over symbolic variables.

\paragraph{GPU programming model.} GPU computation is organized as kernels, where a function is executed in parallel across many threads following the single-program-multiple-data (SPMD) paradigm. Each kernel launch defines a grid of thread blocks, with each block scheduled onto a streaming multiprocessor and containing a group of threads that operate on distinct data elements. Threads maintain private state in registers, while threads within the same block coordinate through low-latency shared memory to support collective operations. Data exchanged between kernels, including inputs and outputs, resides in GPU global memory.

\paragraph{Concrete graph representation.}
Mirage~\cite{mirage} introduces the $\mu$Graph representation for tensor programs, a hierarchical abstraction that specifies tensor programs across the kernel, thread-block, and thread levels of the GPU execution hierarchy. 
In a $\mu$Graph, each graph-defined operator in the kernel graph is associated with a \emph{block graph} (short for thread-block graph) that defines its computation at the block level, and block-level operators may further expand into \emph{thread graphs}. At each level, the graph specifies how input tensors are partitioned via \imap, how output tensors are assembled via \omap, and how loop bodies iterate over reduction dimensions via \fmap. In Mirage, these mappings, along with grid, block, and for-loop dimensions, are instantiated as \emph{concrete} values during search (e.g., $\imap\!: \{r \leftrightarrow x\}$, grid\_dim: $\{x\!=\!64\}$). As a result, each distinct assignment of mappings and dimensions yields a separate $\mu$Graph candidate that must be independently generated, verified, and potentially profiled, leading to a combinatorial explosion that limits search scalability.

\begin{figure*}[t]
    \centering
    \subfloat[Computation Graph]{%
    \begin{tikzpicture}[scale=0.72, every node/.append style={scale=0.72},
        op/.style={rounded corners=2pt, minimum width=1.2cm, minimum height=0.6cm, font=\small\sffamily, fill=knop_color, text=white},
        tensor/.style={minimum width=0.6cm, minimum height=0.6cm, font=\small\sffamily, fill=knop_color, text=white},
        arr/.style={-{Triangle[length=4pt, width=3pt]}, thick},
        every node/.style={font=\small\sffamily},
    ]
    \fill[kngraph_bg]
        (-1.5, -1.2) rectangle (23.0, 1.2);
    \node[font=\footnotesize\sffamily, anchor=north east] at (22.85, 1.15) {\textbf{Kernel Graph}};

    \node[tensor] (A) at (7.5, 0.5) {X};
    \node[font=\scriptsize, anchor=east, xshift=2pt] at (A.west) {\textsf{[4096, 4096]}};
    \node[tensor] (B) at (7.5, -0.5) {W};
    \node[font=\scriptsize, anchor=east, xshift=2pt] at (B.west) {\textsf{[4096, 128]}};
    \node[op] (Softmax) at (9.5, 0.5) {Softmax};
    \node[op] (Matmul) at (12.0, 0) {Matmul};
    \node[tensor] (C) at (13.8, 0) {O};
    \node[font=\scriptsize, anchor=west, xshift=-2pt] at (C.east) {\textsf{[4096, 128]}};
    \draw[arr] (A) -- (Softmax);
    \draw[arr] (Softmax) -- (Matmul);
    \draw[arr] (B) -- (Matmul);
    \draw[arr] (Matmul) -- (C);
    \end{tikzpicture}%
    \label{fig:comp_graph}%
    }%
    \\[0.2em]%
    \subfloat[Concrete $\mu$Graph]{%
    \begin{tikzpicture}[scale=0.72, every node/.append style={scale=0.72},
        op/.style={rounded corners=2pt, minimum width=1.2cm, minimum height=0.6cm, font=\small\sffamily, fill=knop_color, text=white},
        subop/.style={rounded corners=2pt, minimum width=1.2cm, minimum height=0.6cm, font=\small\sffamily, fill=tbop_color, text=white},
        tensor/.style={minimum width=0.6cm, minimum height=0.6cm, font=\small\sffamily, fill=knop_color, text=white},
        arr/.style={-{Triangle[length=4pt, width=3pt]}, thick},
        every node/.style={font=\small\sffamily},
    ]
    \fill[kngraph_bg]
        (-1.5, -3.1) rectangle (4.0, 1.5);
    \node[font=\footnotesize\sffamily, anchor=north east] at (3.85, 1.45) {\textbf{Kernel Graph}};

    \node[tensor] (A) at (-0.5, -0.3) {X};
    \node[font=\scriptsize, anchor=south, yshift=-2pt] at (A.north) {\textsf{[4096, 4096]}};
    \node[tensor] (B) at (-0.5, -1.3) {W};
    \node[font=\scriptsize\sffamily, anchor=north, yshift=2pt] at (B.south) {\textsf{[4096, 128]}};
    \node[op] (CustOp) at (1.25, -0.8) {CustomOp};
    \node[tensor] (C) at (3.0, -0.8) {O};
    \node[font=\scriptsize, anchor=south, yshift=-2pt] at (C.north) {\textsf{[4096, 128]}};
    \draw[arr] (A) -- (CustOp);
    \draw[arr] (B) -- (CustOp);
    \draw[arr] (CustOp) -- (C);

    \draw[dashed, thick, gray] (CustOp.south) -- ++(0, -1.0) -| (8.0, -0.9);

    \fill[tbgraph_bg]
        (4.5, -3.1) rectangle (23.0, 1.5);
    \node[font=\footnotesize\sffamily, anchor=north east] at (22.85, 1.45) {\textbf{ThreadBlock Graph}};
    \node[font=\scriptsize\sffamily, anchor=north west] at (4.65, 1.45) {\textbf{grid\_dim}: \{$x$=64\}\quad \textbf{forloop\_dim}: \{$i$=64\}};

    \node[font=\scriptsize\sffamily, anchor=east, align=left] at (7.3, -0.2) {%
        imap: $\{r\!\leftrightarrow\! x\}$\\fmap: $\{c\!\leftrightarrow\! i\}$};
    \node[font=\scriptsize\sffamily, anchor=east, align=left] at (7.3, -2.0) {%
        imap: $\{\varnothing\}$\\fmap: $\{r\!\leftrightarrow\! i\}$};
    \node[subop, align=center] (In1) at (8.0, -0.2) {Input\\Loader1};
    \node[subop, align=center] (In2) at (8.0, -2.0) {Input\\Loader2};
    \node[font=\scriptsize\sffamily, anchor=south, yshift=-2pt] at (In1.north) {\textsf{[64, 64]}};
    \node[font=\scriptsize\sffamily, anchor=south, yshift=-2pt] at (In2.north) {\textsf{[64, 128]}};
    \node[subop] (Exp) at (10.5, -0.2) {Exp};
    \node[font=\scriptsize\sffamily, anchor=south, yshift=-2pt] at (Exp.north) {\textsf{[64, 64]}};
    \node[subop] (Accum1) at (12.5, -0.2) {Accum};
    \node[font=\scriptsize\sffamily, anchor=south, yshift=-2pt] at (Accum1.north) {\textsf{[64]}};
    \node[subop] (Matmul) at (12.5, -2.0) {Matmul};
    \node[font=\scriptsize\sffamily, anchor=north, yshift=2pt] at (Matmul.south) {\textsf{[64, 128]}};
    \node[subop] (Accum2) at (14.5, -2.0) {Accum};
    \node[font=\scriptsize\sffamily, anchor=north, yshift=2pt] at (Accum2.south) {\textsf{[64, 128]}};
    \node[subop] (Div) at (16.7, -0.9) {Div};
    \node[font=\scriptsize\sffamily, anchor=south, yshift=-2pt] at (Div.north) {\textsf{[64, 128]}};
    \node[subop, align=center] (Out) at (18.9, -0.9) {Output\\Saver};
    \node[font=\scriptsize\sffamily, anchor=south, yshift=-2pt] at (Out.north) {\textsf{[64, 128]}};
    \node[font=\scriptsize\sffamily, anchor=north] at (Out.south) {omap: $\{r\!\leftrightarrow\! x,\; c\!\leftrightarrow\!\varnothing\}$};

    \draw[arr] (In1) -- (Exp);
    \draw[arr] (Exp) -- (Matmul);
    \draw[arr] (In2) -- (Matmul);
    \draw[arr] (Exp) -- (Accum1);
    \draw[arr] (Matmul) -- (Accum2);
    \draw[arr] (Accum1) -- (Div);
    \draw[arr] (Accum2) -- (Div);
    \draw[arr] (Div) -- (Out);
    \end{tikzpicture}%
    \label{fig:symbolic_graph_a}%
    }%
    \\[0.2em]%
    \subfloat[Symbolic graph (sGraph)]{%
    \begin{tikzpicture}[scale=0.72, every node/.append style={scale=0.72},
        op/.style={rounded corners=2pt, minimum width=1.2cm, minimum height=0.6cm, font=\small\sffamily, fill=knop_color, text=white},
        subop/.style={rounded corners=2pt, minimum width=1.2cm, minimum height=0.6cm, font=\small\sffamily, fill=tbop_color, text=white},
        tensor/.style={minimum width=0.6cm, minimum height=0.6cm, font=\small\sffamily, fill=knop_color, text=white},
        arr/.style={-{Triangle[length=4pt, width=3pt]}, thick},
        every node/.style={font=\small\sffamily},
    ]
    \fill[kngraph_bg]
        (-1.5, -4.5) rectangle (4.0, 1.5);
    \node[font=\footnotesize\sffamily, anchor=north east] at (3.85, 1.45) {\textbf{Kernel Graph}};

    \node[tensor] (A) at (-0.5, -1.0) {X};
    \node[font=\scriptsize, anchor=south, yshift=-2pt] at (A.north) {\textsf{[4096, 4096]}};
    \node[tensor] (B) at (-0.5, -2.0) {W};
    \node[font=\scriptsize\sffamily, anchor=north, yshift=2pt] at (B.south) {\textsf{[4096, 128]}};
    \node[op] (CustOp) at (1.25, -1.5) {CustomOp};
    \node[tensor] (C) at (3.0, -1.5) {O};
    \node[font=\scriptsize, anchor=south, yshift=-2pt] at (C.north) {\textsf{[4096, 128]}};
    \draw[arr] (A) -- (CustOp);
    \draw[arr] (B) -- (CustOp);
    \draw[arr] (CustOp) -- (C);

    \draw[dashed, thick, gray] (CustOp.south) -- ++(0, -0.8) -| (8.0, -1.0);

    \fill[tbgraph_bg]
        (4.5, -4.5) rectangle (23.0, 1.5);
    \node[font=\footnotesize\sffamily, anchor=north east] at (22.85, 1.45) {\textbf{ThreadBlock Graph}};
    \node[font=\scriptsize\sffamily, anchor=north west] at (4.65, 1.45) {\textbf{grid\_dim}: \{$x$=$d_x$\}\quad \textbf{forloop\_dim}: \{$i$=$d_i$\}};

    \newcommand{\sfrac}[3]{{\tfrac{#1}{\sigma(#2,#3)}}}
    \node[font=\scriptsize\sffamily, anchor=east, align=left] at (7.3, -0.9) {%
        imap: $\left[\begin{smallmatrix} m_{X,r,x} \\[1pt] m_{X,c,x} \end{smallmatrix}\right]$\\
        fmap: $\left[\begin{smallmatrix} m_{X,r,i} \\[1pt] m_{X,c,i} \end{smallmatrix}\right]$};
    \node[font=\scriptsize\sffamily, anchor=east, align=left] at (7.3, -2.7) {%
        imap: $\left[\begin{smallmatrix} m_{W,r,x} \\[1pt] m_{W,c,x} \end{smallmatrix}\right]$\\
        fmap: $\left[\begin{smallmatrix} m_{W,r,i} \\[1pt] m_{W,c,i} \end{smallmatrix}\right]$};

    \node[subop, align=center] (In1) at (8.0, -0.9) {Input\\Loader1};
    \node[subop, align=center] (In2) at (8.0, -2.7) {Input\\Loader2};
    \node[font=\scriptsize\sffamily, anchor=south, yshift=-1pt] at (In1.north) {$\left[\sfrac{4096}{X}{r},\; \sfrac{4096}{X}{c}\right]$};
    \node[font=\scriptsize\sffamily, anchor=north, yshift=1pt] at (In2.south) {$\left[\sfrac{4096}{W}{r},\; \sfrac{128}{W}{c}\right]$};
    \node[subop] (Exp) at (10.5, -0.9) {Exp};
    \node[font=\scriptsize\sffamily, anchor=south, yshift=-1pt] at (Exp.north) {$\left[\sfrac{4096}{X}{r},\; \sfrac{4096}{X}{c}\right]$};
    \node[subop] (Accum1) at (12.5, -0.9) {Accum};
    \node[font=\scriptsize\sffamily, anchor=south, xshift=8pt, yshift=-1pt] at (Accum1.north) {$\left[\sfrac{4096}{X}{r}\right]$};
    \node[subop] (Matmul) at (12.5, -2.7) {Matmul};
    \node[font=\scriptsize\sffamily, anchor=north, xshift=-6pt, yshift=1pt] at (Matmul.south) {$\left[\sfrac{4096}{X}{r},\; \sfrac{128}{W}{c}\right]$};
    \node[subop] (Accum2) at (14.5, -2.7) {Accum};
    \node[font=\scriptsize\sffamily, anchor=north, xshift=6pt, yshift=1pt] at (Accum2.south) {$\left[\sfrac{4096}{X}{r},\; \sfrac{128}{W}{c}\right]$};
    \node[subop] (Div) at (16.7, -1.6) {Div};
    \node[font=\scriptsize\sffamily, anchor=south, yshift=-1pt] at (Div.north) {$\left[\sfrac{4096}{X}{r},\; \sfrac{128}{W}{c}\right]$};
    \node[subop, align=center] (Out) at (19.5, -1.6) {Output\\Saver};
    \node[font=\scriptsize\sffamily, anchor=south, yshift=-1pt] at (Out.north) {$\left[\sfrac{4096}{O}{r},\; \sfrac{128}{O}{c}\right]$};
    \node[font=\scriptsize\sffamily, anchor=west, align=left] at (20.3, -1.6) {%
        omap: $\left[\begin{smallmatrix} m_{O,r,x} \\[1pt] m_{O,c,x} \end{smallmatrix}\right]$};

    \draw[arr] (In1) -- (Exp);
    \draw[arr] (Exp) -- (Matmul);
    \draw[arr] (In2) -- (Matmul);
    \draw[arr] (Exp) -- (Accum1);
    \draw[arr] (Matmul) -- (Accum2);
    \draw[arr] (Accum1) -- (Div);
    \draw[arr] (Accum2) -- (Div);
    \draw[arr] (Div) -- (Out);
    \node[font=\scriptsize, anchor=south east, align=left] at (22.8, -4.4) {%
        $\sigma(T,d)\!=\!\prod_{p \in \mathcal{P}}(m_{T,d,p}\!\cdot\! d_p + 1 - m_{T,d,p})$};
    \end{tikzpicture}%
    \label{fig:symbolic_graph_b}%
    }%
    \caption{Graph representations of a fused Softmax-Matmul operation. (a) The input computation graph. (b) A concrete $\mu$Graph with specific mappings and parallelization parameters. (c) Our symbolic graph (\graph), where mappings and dimensions are represented as symbolic variables.}
    \label{fig:symbolic_graph}
\end{figure*}
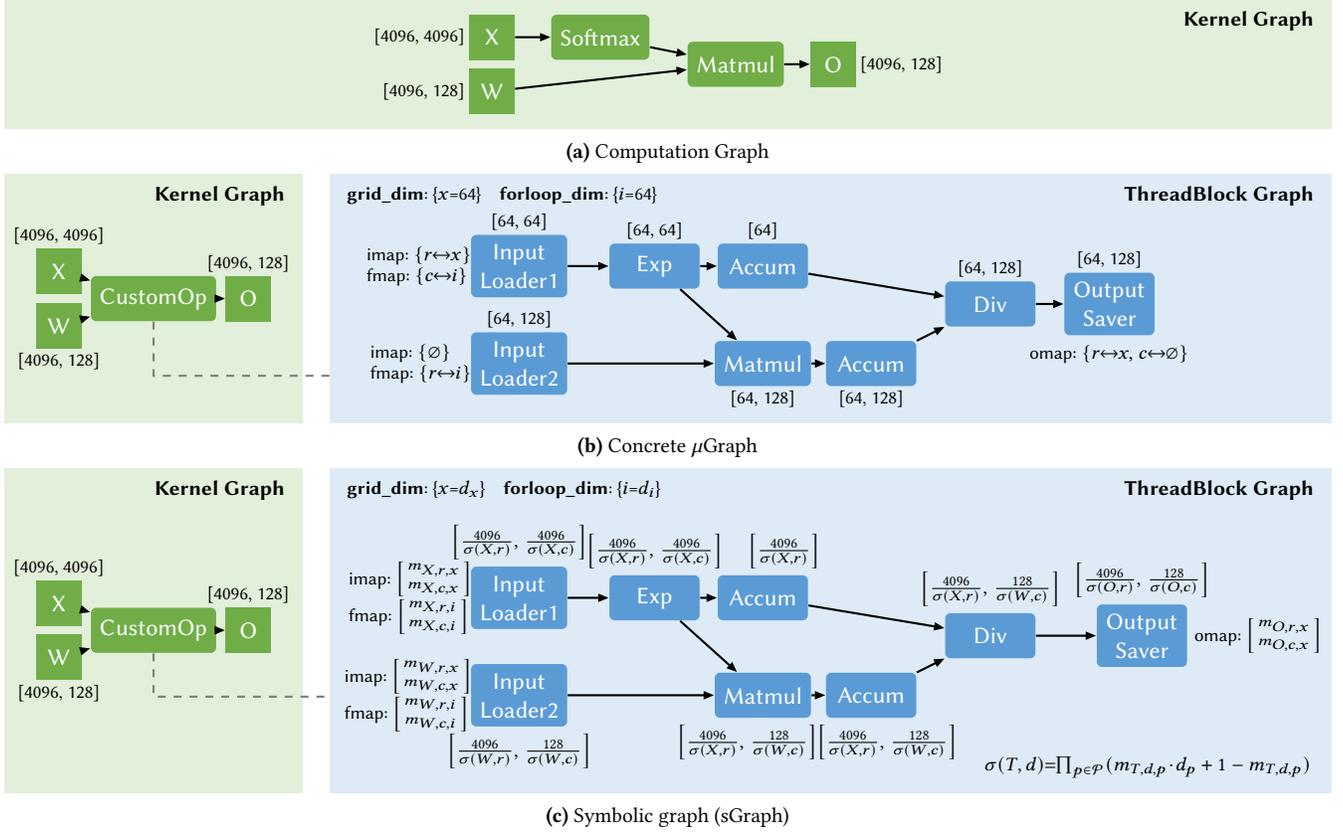

\paragraph{Symbolic graph representation.} \Sys introduces \emph{symbolic graphs} (\graphs), which generalize $\mu$Graphs by replacing concrete dimensions and mappings with symbolic variables. An \graph retains the same hierarchical structure---a kernel graph whose operators expand into block graphs, which may further expand into thread graphs---but represents grid and block dimensions using symbolic integer variables and mappings (\imap, \fmap, \omap) using symbolic Boolean variables. As a result, a single \graph compactly represents a family of $\mu$Graphs parameterized by the variables, enabling symbolic reasoning over the entire family without explicitly enumerating individual candidates. Figure~\ref{fig:symbolic_graph} illustrates this process:
(a) shows the computation graph of a Softmax followed by a matrix multiplication, 
(b) shows a concrete $\mu$Graph for a fused kernel with fixed mappings and dimensions, 
and (c) shows the corresponding \graph where the same structure is expressed symbolically.

\paragraph{Symbolic parallelization parameters.}
\Cref{fig:symbolic_graph_a} shows a concrete $\mu$Graph, where grid dimensions (e.g., $x = 64$), block dimensions, and for-loop dimensions (e.g., $i = 64$) are fixed integers that determine how computation is distributed across parallel execution units. 
We collectively refer to these as \emph{parallelization dimensions}. For a block graph, we denote the set of parallelization dimensions as $\mathcal{P} = \mathcal{P}_g \cup \mathcal{P}_f$, where $\mathcal{P}_g$ and $\mathcal{P}_f$ are the sets of grid and for-loop dimensions, respectively. 
The same formalism applies to thread graphs, where $\mathcal{P}_g$ consists of block dimensions and $\mathcal{P}_f$ is empty. Each parallelization dimension $p \in \mathcal{P}$ has an associated size $d_p$, and we denote the vector of all sizes as $\mathbf{d} = (d_p)_{p \in \mathcal{P}}$. 

In an \graph, these sizes are left symbolic, allowing a single \graph to represent a family of $\mu$Graphs with different parallelization granularities.
Currently, we assume a single for-loop dimension (i.e., so $|\mathcal{P}_f|=1$), which is sufficient for the optimizations we consider; 
extending the framework to multiple loop dimensions is straightforward.

\paragraph{Symbolic mappings.}
At each level of a $\mu$Graph, the mappings (i.e., \imap, \fmap, \omap) specify how tensors are partitioned or replicated across parallelization dimensions. In an \graph, we symbolically encode these mappings using Boolean variables. For each tensor $T$ and each pair of data dimension $d$ and parallelization dimension $p \in \mathcal{P}$, we introduce a variable $m_{T, d, p} \in \{0, 1\}$, where $m_{T,d,p} = 1$ indicates that 
$d$ is partitioned along $p$.
If a parallelization dimension $p$ does not partition any data dimension of $T$, then $T$ is replicated along $p$.
For input tensors, these variables encode the \imap and \fmap mappings. For output tensors, only grid dimensions appear in \omap, as reductions along loop dimensions are handled explicitly by accumulator operators.

These variables must satisfy two families of constraints:
\begin{align}
    &\textstyle\sum_{d \in \text{dims}(T)} m_{T, d, p} \le 1, \quad \forall\, p \in \mathcal{P}
    \label{eq:constraint_parallel} \\
    &\textstyle\sum_{p \in \mathcal{P}_g} m_{T, d, p} \le 1, \quad \forall\, d \in \text{dims}(T)
    \label{eq:constraint_data}
\end{align}
Constraint~\eqref{eq:constraint_parallel} requires that each parallelization dimension partitions at most one data dimension of $T$. Constraint~\eqref{eq:constraint_data} requires that each data dimension is partitioned by at most one grid dimension in $\mathcal{P}_g$.
Note that a corresponding constraint is unnecessary for $\mathcal{P}_f$ since we assume a single for-loop dimension.
A grid dimension in $\mathcal{P}_g$ and the for-loop dimension in $\mathcal{P}_f$ may partition the same data dimension, as they operate independently.

Figure~\ref{fig:symbolic_graph}(c) illustrates a block graph with one grid dimension $x$ and one for-loop dimension $i$, where each tensor has a row dimension $r$ and column dimension $c$. In this example, the \imap and \fmap of InputLoader1 (tensor $X$) are encoded as $[m_{X,r,x},\; m_{X,c,x}]$ and $[m_{X,r,i},\; m_{X,c,i}]$, respectively. The concrete $\mu$Graph in Figure~\ref{fig:symbolic_graph}(b), where $\imap:{r \leftrightarrow x}$ and $\fmap:{c \leftrightarrow i}$, corresponds to the assignment $m_{X,r,x} = 1$, $m_{X,c,x} = 0$, $m_{X,r,i} = 0$, and $m_{X,c,i} = 1$. Replication is represented by all mapping variables being zero for a given parallelization dimension; for example, InputLoader2 (tensor $W$) has $\imap:{\varnothing}$, corresponding to $m_{W,r,x} = m_{W,c,x} = 0$.

\paragraph{Symbolic tensor shapes.}
Given symbolic mappings and symbolic parallelization parameters, tensor shapes become symbolic expressions. Consider an input tensor $T$ of a block graph with a data dimension $d$ of original size $D$. The per-block, per-iteration size of $d$ is:
\begin{align}
    \frac{D}{\sigma(T, d)}, \quad \text{where} \quad
    \sigma(T, d) = \textstyle\prod_{p \in \mathcal{P}} \left( m_{T,d,p} \cdot d_p + 1 - m_{T,d,p} \right)
    \label{eq:symbolic_dim}
\end{align}
Each factor in the product evaluates to $d_p$ when $m_{T,d,p} = 1$ (i.e., dimension $d$ is partitioned along $p$) and to $1$ otherwise. Since Constraint~\eqref{eq:constraint_data} ensures that at most one grid dimension in $\mathcal{P}_g$ and at most one for-loop dimension partition $d$, the product reduces to the sizes of the active parallelization dimensions. The shapes of intermediate tensors are then derived from operator semantics (e.g., a matrix multiplication of tensors with shapes $[a, b]$ and $[b, c]$ produces a tensor of shape $[a, c]$).

For example, in Figure~\ref{fig:symbolic_graph}(c), the block graph has one grid dimension $x$ and one loop dimension $i$, so $\sigma(T,d) = (m_{T,d,x} \cdot d_x + 1 - m_{T,d,x}) \cdot (m_{T,d,i} \cdot d_i + 1 - m_{T,d,i})$. The shape of InputLoader1 (tensor $X$ with original shape $[4096, 4096]$) in the block graph is $[\tfrac{4096}{\sigma(X,r)},\; \tfrac{4096}{\sigma(X,c)}]$. Under the concrete assignment in Figure~\ref{fig:symbolic_graph}(b), where $m_{X,r,x} = 1$ and $m_{X,c,i} = 1$, we get $\sigma(X,r) = d_x \cdot 1 = 64$ and $\sigma(X,c) = 1 \cdot d_i = 64$, so the shape evaluates to $[64, 64]$.

\paragraph{Symbolic shape matching.}
For a $\mu$Graph to be valid, tensor shapes at each operator must be compatible---for example, a matrix multiplication requires matching contracting dimensions. In an \graph, tensor shapes are symbolic expressions over mapping variables $\mathbf{m}$ and parallelization parameters $\mathbf{d}$, therefore shape compatibility is enforced through constraints over these symbolic variables.

As an example, consider the Matmul operator in Figure~\ref{fig:symbolic_graph}(c), which multiplies the output of Exp (shape inherited from InputLoader1) with InputLoader2 (tensor $W$, original shape $[4096, 128]$). For the contracting dimensions to match, the $c$-dimension of Exp must be equivalent to the $r$-dimension of InputLoader2:
\begin{align}
    \tfrac{4096}{\sigma(X,c)} = \tfrac{4096}{\sigma(W,r)}
    \label{eq:shape_match}
\end{align}
This equality holds when both dimensions are partitioned identically, i.e., $m_{X,c,p} = m_{W,r,p}$ for all $p \in \mathcal{P}$. Together with the linear constraints from Equations~\eqref{eq:constraint_parallel}--\eqref{eq:constraint_data} (e.g., $m_{X,r,x} + m_{X,c,x} \le 1$, ensuring at most one dimension of $X$ is partitioned along $x$), these shape-matching constraints restrict the space of valid assignments to the symbolic mappings and parallelization parameters for a given \graph.

\paragraph{Correct mappings and feasible \graphs.}
Given the symbolic representation above, we formalize when an \graph correctly implements a given input program. A key design choice is to require correctness to hold for \emph{all} values of the parallelization parameters, rather than for specific assignments. This requirement ensures that a mapping remains correct regardless of the chosen parallelization granularity, which is important because different hardware configurations or input sizes may call for different parameter values. It also cleanly decouples correctness verification from parameter tuning: once a mapping is verified to be correct, the parallelization parameters can be tuned for performance without re-validating equivalence.

\begin{definition}[Correct Mapping]
    Given an input tensor program $G_{\mathrm{in}}$ and an \graph $S$, a \emph{correct mapping} is an assignment $\hat{\mathbf{m}}$ of the mapping variables $\mathbf{m}$ such that for every assignment $\hat{\mathbf{d}}$ of the parallelization parameters $\mathbf{d}$,  the instantiated graph $S_{\hat{\mathbf{m}},\, \hat{\mathbf{d}}}$ computes the same function as $G_{\mathrm{in}}$.
\end{definition}

\begin{definition}[Feasible \graph]
    Given an input tensor program $G_{\mathrm{in}}$, a \emph{feasible \graph} is an \graph that has at least one correct mapping.
\end{definition}

%% file: generate.tex
\section{\graph Generation}
\label{sec:search}

\subsection{Generator Overview}

\begin{figure*}[t]
    \centering
    \begin{tikzpicture}[
        scale=0.55, every node/.append style={scale=0.55},
        gbox/.style={draw=none, rounded corners=3pt, fill=kngraph_bg, minimum width=2.0cm, minimum height=0.55cm, font=\normalsize\sffamily, inner sep=3pt, align=center},
        arr/.style={-{Triangle[length=3pt, width=2pt]}, thick, pipearrow},
        fatarr/.style={-{Triangle[length=4pt, width=3pt]}, very thick, pipearrow},
        every node/.style={font=\normalsize\sffamily},
        checkmark/.style={font=\normalsize, color=green!60!black},
        crossmark/.style={font=\normalsize, color=red!70!black},
    ]
    \def\stripW{28}
    \def\stripH{4.6}
    \def\arrowX{31.5}

    \def\yA{0}
    \def\stripHA{4.0}
    \fill[pipebg] (0, \yA+0.6) rectangle (\stripW, \yA-\stripHA);
    \node[font=\normalsize\sffamily\bfseries, anchor=north west] at (0.3, \yA+0.5) {\graph Generation (\S\ref{sec:search})};

    \node[font=\small] at (\stripW/2, \yA+0.3) {$\cdots$};
    \draw[fatarr] (\stripW/2, \yA+0.1) -- (\stripW/2, \yA-0.5);
    \node[gbox, minimum width=2.5cm] (G) at (\stripW/2, \yA-0.9) {Partial \graph $G$};

    \node[gbox] (Gop1) at (6, \yA-2.6) {$G$ + op$_1$};
    \node[gbox] (Gop2) at (14, \yA-2.6) {$G$ + op$_2$};
    \node[gbox] (Gop3) at (22, \yA-2.6) {$G$ + op$_3$};
    \draw[fatarr] (G.south) -- (Gop1.north);
    \draw[fatarr] (G.south) -- (Gop2.north);
    \draw[fatarr] (G.south) -- (Gop3.north);

    \node[font=\small\sffamily, text=red!70!black] at (6, \yA-3.5) {Dim mismatch \texttimes};

    \node[font=\small\sffamily, text=red!70!black] at (14, \yA-3.5) {Not subexpr \texttimes};

    \node[font=\small\sffamily, text=green!60!black] at (22, \yA-3.5) {Dim match, expr check \checkmark};
    \draw[fatarr] (Gop3.east) -- ++(1.5, 0) node[right, font=\small] {$\cdots$};

    \draw[{Triangle[length=6pt, width=5pt]}-, ultra thick, black] (-0.3, \yA-\stripHA/2+0.3) -- (-3.0, \yA-\stripHA/2+0.3)
        node[midway, above, font=\small\sffamily, align=center] {Input tensor\\program};

    \def\yB{-5.0}
    \fill[pipebg] (0, \yB+0.6) rectangle (\stripW, \yB-\stripH);
    \node[font=\normalsize\sffamily\bfseries, anchor=north west] at (0.3, \yB+0.5) {Mapping Instantiation (\S\ref{sec:search:instantiation})};

    \node[font=\small\sffamily\bfseries, anchor=north west] at (0.3, \yB-0.5) {Symbolic mappings:};
    \node[font=\small, anchor=north west] at (0.3, \yB-1.1) {$m_X\!=\!\left[\begin{smallmatrix} m_{X,r,x} & m_{X,r,i} \\ m_{X,c,x} & m_{X,c,i} \end{smallmatrix}\right]$};
    \node[font=\small, anchor=north west] at (0.3, \yB-1.9) {$m_W\!=\!\left[\begin{smallmatrix} m_{W,r,x} & m_{W,r,i} \\ m_{W,c,x} & m_{W,c,i} \end{smallmatrix}\right]$};
    \node[font=\small, anchor=north west] at (0.3, \yB-2.7) {$m_O\!=\!\left[\begin{smallmatrix} m_{O,r,x} \\ m_{O,c,x} \end{smallmatrix}\right]$};

    \node[font=\small\sffamily\bfseries, anchor=north west] at (6.0, \yB-0.5) {Constraints:};
    \node[font=\small, anchor=north west, align=left] at (6.0, \yB-1.1) {%
        (1) \textsf{mapping def.:} $m_{X,r,x}\!+\!m_{X,c,x}\!\le\!1$, \ldots\\
        (2) \textsf{dim match:} $m_{X,c,x}\!=\!m_{W,r,x}$,\; $m_{X,c,i}\!=\!m_{W,r,i}$\\
        (3) \textsf{symmetry:} lexicographically smallest};

    \node[font=\small\sffamily\bfseries, anchor=north west] at (17, \yB-0.5) {Enumerate:};

    \node[font=\small, anchor=north west, fill=pipepass, rounded corners=1pt, inner sep=2pt] at (17, \yB-1.3) {%
        $\hat{m}_X\!=\!\left[\begin{smallmatrix} 1 & 0 \\ 0 & 1 \end{smallmatrix}\right]$, $\hat{m}_W\!=\!\left[\begin{smallmatrix} 0 & 1 \\ 0 & 0 \end{smallmatrix}\right]$, $\hat{m}_O\!=\!\left[\begin{smallmatrix} 1 \\ 0 \end{smallmatrix}\right]$ \checkmark};

    \node[font=\small, anchor=north west, fill=pipefail, rounded corners=1pt, inner sep=2pt] at (17, \yB-2.3) {%
        $\hat{m}_X\!=\!\left[\begin{smallmatrix} 1 & 0 \\ 1 & 0 \end{smallmatrix}\right]$, $\hat{m}_W\!=\!\left[\begin{smallmatrix} 1 & 0 \\ 0 & 0 \end{smallmatrix}\right]$, $\hat{m}_O\!=\!\left[\begin{smallmatrix} 1 \\ 0 \end{smallmatrix}\right]$ \texttimes\ violates (1)};

    \node[font=\small, anchor=north west, fill=pipefail, rounded corners=1pt, inner sep=2pt] at (17, \yB-3.3) {%
        $\hat{m}_X\!=\!\left[\begin{smallmatrix} 1 & 0 \\ 0 & 1 \end{smallmatrix}\right]$, $\hat{m}_W\!=\!\left[\begin{smallmatrix} 1 & 1 \\ 0 & 0 \end{smallmatrix}\right]$, $\hat{m}_O\!=\!\left[\begin{smallmatrix} 1 \\ 0 \end{smallmatrix}\right]$ \texttimes\ violates (2)};

    \draw[{Triangle[length=6pt, width=5pt]}-, ultra thick, black] (-0.3, \yB-\stripH/2+0.3) -- (-3.0, \yB-\stripH/2+0.3)
        node[midway, above, font=\small\sffamily, align=center] {\graphs};

    \def\yC{-10.6}
    \def\stripHC{3.6}
    \fill[pipebg] (0, \yC+0.6) rectangle (\stripW, \yC-\stripHC);
    \node[font=\normalsize\sffamily\bfseries, anchor=north west] at (0.3, \yC+0.5) {\graph Verification (\S\ref{sec:optimizer})};

    \node[font=\small\sffamily\bfseries, anchor=north west] at (0.5, \yC-0.5) {Input:};
    \node[font=\small, anchor=north west] at (5.0, \yC-0.5) {$E_{\mathrm{input}} = \mathsf{matmul}(\mathsf{div}(\mathsf{exp}(v_X),\; \mathsf{sum}(\mathsf{exp}(v_X))),\; v_W)$};

    \node[font=\small\sffamily\bfseries, anchor=north west] at (0.5, \yC-1.2) {Candidate:};
    \node[font=\small, anchor=north west] at (5.0, \yC-1.2) {$E_{\mathrm{cand}} = \mathsf{comb}(\mathsf{div}(\mathsf{matmul}(\mathsf{exp}(\mathsf{part}(v_X, r, x)),\, \mathsf{part}(v_W, c, x)),\; \mathsf{red}(\mathsf{exp}(\mathsf{part}(v_X, r, x)), x)),\; r, x)$};

    \node[font=\small\sffamily, anchor=north west] at (2.0, \yC-2.2) {Check $\mathsf{equivalent}(E_{\mathrm{input}},\, E_{\mathrm{cand}})$ using axioms in Table~\ref{tab:rewriting_rules}};
    \node[checkmark, font=\normalsize\sffamily\bfseries, anchor=north west] at (20.0, \yC-2.2) {\checkmark\ Equivalent $\Rightarrow$ Verified};

    \draw[{Triangle[length=6pt, width=5pt]}-, ultra thick, black] (-0.3, \yC-\stripHC/2+0.3) -- (-3.0, \yC-\stripHC/2+0.3)
        node[midway, above, font=\small\sffamily, align=center] {\graphs with\\concrete $\hat{\mathbf{m}}$};

    \def\yD{-15.2}
    \def\stripHD{3.5}
    \fill[pipebg] (0, \yD+0.6) rectangle (\stripW, \yD-\stripHD);
    \node[font=\normalsize\sffamily\bfseries, anchor=north west] at (0.3, \yD+0.5) {Parameter Instantiation (\S\ref{sec:instantiation})};

    \node[gbox, minimum width=1.6cm, font=\normalsize\sffamily\bfseries] (sym) at (3.0, \yD-1.8) {Symbolic\\graph};
    \node[gbox, minimum width=1.6cm, font=\normalsize\sffamily\bfseries] (con) at (13.0, \yD-1.8) {Concrete\\graph};
    \draw[-{Triangle[length=6pt, width=5pt]}, very thick, pipearrow] (sym.east) -- (con.west)
        node[midway, above, font=\small\sffamily] {Random sample}
        node[midway, below, font=\small] {$\mathbf{d}\!=\!(d_x, d_i)$};

    \node[font=\small\sffamily\bfseries, anchor=north west] at (16, \yD+0.4) {Profile:};
    \node[font=\small, anchor=north west, fill=pipebox!30, rounded corners=1pt, inner sep=2pt] at (16, \yD-0.3) {$d_x\!=\!32,\; d_i\!=\!64$ $\to$ 0.089ms};
    \node[font=\small, anchor=north west, fill=pipebox!30, rounded corners=1pt, inner sep=2pt] at (16, \yD-1.1) {$d_x\!=\!64,\; d_i\!=\!64$ $\to$ 0.059ms};
    \node[font=\small, anchor=north west, fill=pipepass, rounded corners=1pt, inner sep=2pt] (best) at (16, \yD-1.9) {$d_x\!=\!64,\; d_i\!=\!32$ $\to$ \textbf{0.042ms} \checkmark};
    \node[font=\small\sffamily\bfseries, anchor=west, color=pipearrow, align=left] at (best.east) {\,$\Rightarrow$ Optimized\\\ \,\ \ tensor program};
    \node[font=\small, anchor=north west, fill=pipebox!30, rounded corners=1pt, inner sep=2pt] at (16, \yD-2.7) {$d_x\!=\!128,\; d_i\!=\!32$ $\to$ 0.056ms};

    \draw[{Triangle[length=6pt, width=5pt]}-, ultra thick, black] (-0.3, \yD-\stripHD/2+0.3) -- (-3.0, \yD-\stripHD/2+0.3)
        node[midway, above, font=\small\sffamily, align=center] {Verified \graphs\\with concrete $\hat{\mathbf{m}}$};

    \end{tikzpicture}
    \caption{Overview of the \sys pipeline. \textbf{\graph Generation} (\S\ref{sec:search}): exhaustive search builds \graphs with symbolic mappings; dimension matching and expression-guided pruning eliminate invalid branches. \textbf{Mapping Instantiation} (\S\ref{sec:search:instantiation}): enumerates candidate concrete mapping assignments satisfying all constraints. \textbf{\graph Verification} (\S\ref{sec:optimizer}): equivalence checking using rewrite axioms. \textbf{Parameter Instantiation} (\S\ref{sec:instantiation}): random sampling with GPU profiling tunes parallelization parameters.}
    \label{fig:pipeline}
\end{figure*}
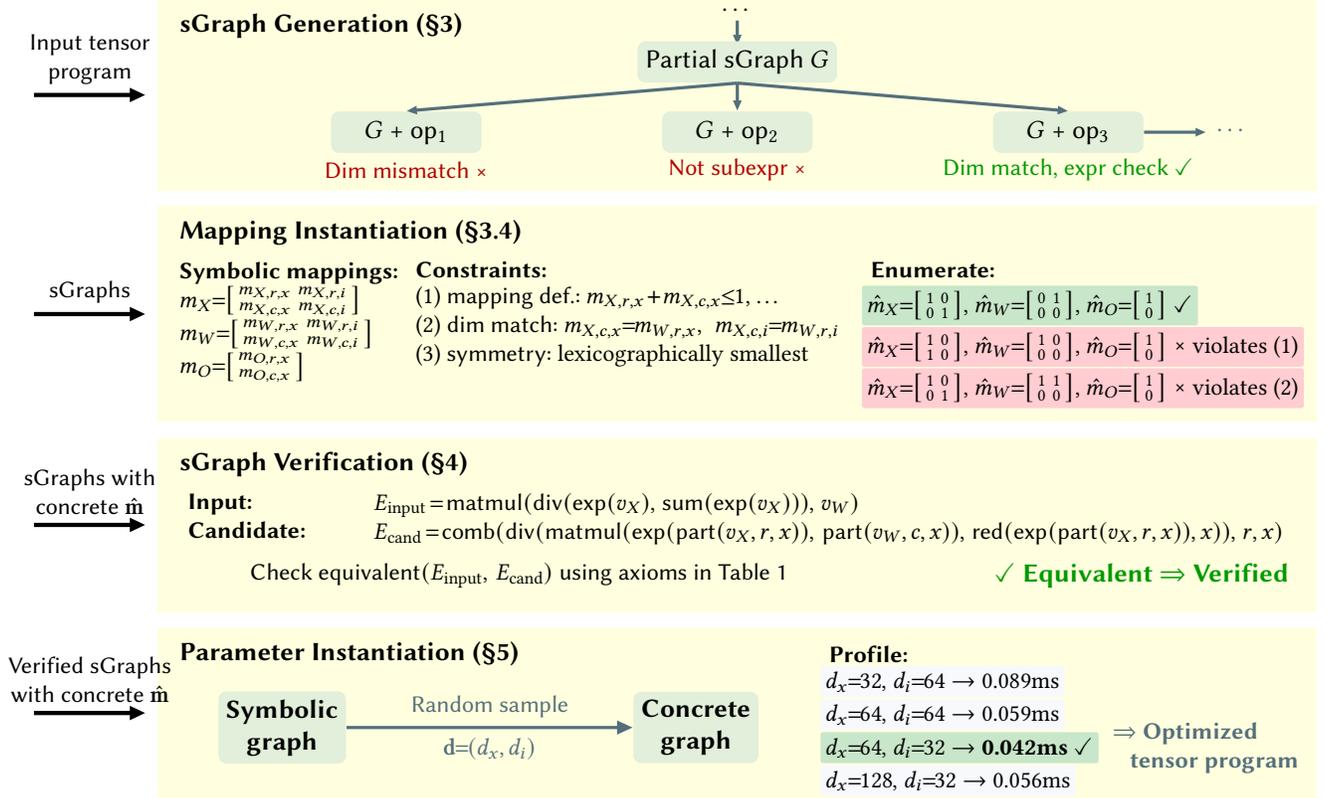

The goal of the \graph generator is to efficiently search for feasible \graphs and their corresponding correct mappings for a given input program. The generator uses an iterative approach similar to prior work: it constructs candidate graphs by incrementally adding operators and checking validity at each step. 

Unlike prior approaches that enumerate concrete mapping assignments during graph construction, the \graph generator introduces symbolic mapping variables (\S\ref{sec:hcg}) and performs symbolic shape matching to determine whether an operator can be validly added. This design allows the generator to explore the space of graph structures without committing to specific mappings, thereby avoiding early combinatorial blowup. The enumeration of concrete mappings is deferred to a later verification stage, while parallelization parameters are instantiated only after correctness has been established (\S\ref{sec:search:instantiation}).

\paragraph{Search space comparison.}
In concrete superoptimizers, the generator must enumerate all combinations of graph structures, concrete mappings, and parallelization parameter values, leading to a search space of $O(|\mathcal{G}||\mathcal{M}||\mathcal{D}|)$ where $|\mathcal{G}|$ is the number of graph structures, $|\mathcal{M}|$ is the average number of mapping assignments per graph, and $|\mathcal{D}|$ is the average number of parallelization parameter configurations.

In contrast, the \graph generator decouples structure search from both mapping enumeration and parameter tuning. It explores graph structures once with symbolic mappings and symbolic parallelization parameters ($O(|\mathcal{G}|)$), prunes invalid candidates via expression-guided pruning (\S\ref{sec:search:pruning}), and defers the enumeration of concrete mappings to a mapping instantiation phase (\S\ref{sec:search:instantiation}) and the tuning of parallelization parameters to a subsequent parameter instantiation phase (\S\ref{sec:instantiation}). Since pruning eliminates a large fraction of candidates early, the overall search cost is significantly reduced.

\subsection{Symbolic Dimension Matching}
\label{sec:search:dim_matching}

As described in \S\ref{sec:hcg}, shape compatibility in an \graph imposes constraints on both the mapping variables $\mathbf{m}$ and the parallelization parameters $\mathbf{d}$ (Equation~\ref{eq:shape_match}). By the definition of correct mapping, shape compatibility must hold for {\em all} values of $\mathbf{d}$. Therefore, the matched dimension expressions must be identical as functions of $\mathbf{d}$, reducing shape matching to constraints purely over $\mathbf{m}$. In particular, we equate the mapping variables so that the two symbolic expressions become identical with respect to $\mathbf{d}$.

When adding an operator to a partial \graph, the generator performs two tasks: (1) it collects equality constraints over the mapping variables, which are enforced later during mapping instantiation (\S\ref{sec:search:instantiation}), and (2) it checks whether the resulting dimension expressions are compatible, immediately pruning partial \graphs that fail this check.

As a concrete example, consider adding the Matmul operator in Figure~\ref{fig:symbolic_graph}(c). Its contracting dimensions must match: the column dimension of the left input (from Exp, inherited from InputLoader1) has symbolic size $\tfrac{4096}{\sigma(X,c)}$, and the row dimension of the right input (InputLoader2 for tensor $W$) has symbolic size $\tfrac{4096}{\sigma(W,r)}$.

\paragraph{Equality constraints on mapping variables.}
In general, identifying the appropriate mapping variables to equate can be complex for arbitrary expressions. In our setting, however, symbolic dimension expressions are sufficiently structured that coefficient matching is effective. Specifically, we group mapping variables by the parallelization parameters they are associated with and equate those that appear as coefficients of the same parameter. In the Matmul example, $\sigma(X,c)$ contains terms $m_{X,c,x} \cdot d_x$ and $m_{X,c,i} \cdot d_i$, while $\sigma(W,r)$ contains $m_{W,r,x} \cdot d_x$ and $m_{W,r,i} \cdot d_i$. Matching coefficients of $d_x$ and $d_i$ yields the constraints $m_{X,c,x} = m_{W,r,x}$ and $m_{X,c,i} = m_{W,r,i}$.

\paragraph{Compatibility check.}
After applying these equality constraints, we check whether the resulting dimension expressions, composed of basic arithmetic operations ($+$, $-$, $\times$, $\div$), are symbolically equivalent. If they are not, the operator is deemed incompatible and the partial \graph is pruned.

\subsection{Expression-Guided Pruning}
\label{sec:search:pruning}

In symbolic graphs, tensor shapes and expressions depend on both the mapping variables $\mathbf{m}$ and the parallelization parameters $\mathbf{d}$, so standard expression-based pruning---which operates on concrete tensor shapes---cannot be applied directly.

Our key observation is that any completion of a partial \graph into a feasible \graph must satisfy the expression check for \emph{all} values of $\hat{\mathbf{d}}$. Therefore, checking the condition under a single concrete assignment yields a necessary condition for feasibility. We choose $\hat{\mathbf{d}} = \mathbf{1}$, under which $\sigma(T,d) = 1$ for all tensors and dimensions, making tensor shapes independent of $\mathbf{m}$. This reduces the partial \graph to a non-symbolic graph with concrete tensor shapes, to which we apply the abstract expression checking from Mirage~\cite{mirage}: we check whether the abstract expression of each intermediate tensor is a subexpression of the final output expression, and prune partial graphs that fail this condition.

By design, this check is \emph{under-pruning}: it never discards a partial graph that could lead to a feasible \graph, but may retain some infeasible candidates, which are subsequently filtered out during mapping instantiation and verification (\S\ref{sec:search:instantiation}, \S\ref{sec:optimizer}). In practice, this inexpensive check prunes a substantial portion of the search space.

\subsection{Mapping Instantiation}
\label{sec:search:instantiation}

After generating candidate \graphs that pass the pruning check, we \emph{partially} instantiate each \graph by enumerating assignments to the mapping variables $\mathbf{m}$, while keeping the parallelization parameters $\mathbf{d}$ symbolic (to be tuned in a later phase). For each candidate mapping, we verify correctness using the e-graph-based equivalence checking from \S\ref{sec:optimizer}.

\paragraph{Enumerating candidate mappings.}
A valid mapping must satisfy two classes of constraints:
\begin{compactitem}
\item \textbf{Linear constraints} (Equations~\eqref{eq:constraint_parallel}--\eqref{eq:constraint_data}): each parallelization dimension maps to at most one data dimension, and each data dimension is mapped to at most one grid dimension in $\mathcal{P}_g$.
\item \textbf{Equality constraints} from symbolic dimension matching (\S\ref{sec:search:dim_matching}): matched dimensions must be partitioned identically.
\end{compactitem}
We enumerate candidate mappings by exploring all combinations of \imap, \fmap, and \omap, and filtering out those that violate any of these constraints.

\paragraph{Symmetry breaking.}
Different mapping assignments may yield functionally identical $\mu$Graphs when they differ only by a permutation of dimensions in $\mathcal{P}_g$. To eliminate redundant verification, \sys retains only the lexicographically smallest assignment within each equivalence class, reducing the number of candidates by up to a factor of $k!$ for $k = |\mathcal{P}_g|$.

%% file: verification.tex
\section{\graph Verification}
\label{sec:optimizer}

After mapping instantiation (\S\ref{sec:search:instantiation}), each candidate \graph has concrete mappings but symbolic parallelization parameters. To verify that such a partially-instantiated \graph is functionally equivalent to the input computation graph, we encode both as expressions and check their equivalence. We define a set of equivalence axioms (Table~\ref{tab:rewriting_rules}) that capture the mathematical properties of the operators, and use e-graphs~\cite{egg} to check whether two expressions are equivalent under these axioms. This section describes the expression language and the axioms; the practical details of converting axioms to e-graph rewrite rules are discussed in \S\ref{sec:impl}.

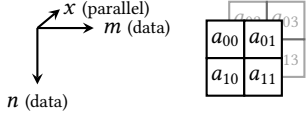
\begin{figure}
    \centering
    \begin{tikzpicture}[
        cell/.style={minimum size=0.5cm, draw, thick, inner sep=0pt, font=\small},
        dimlbl/.style={font=\small, ->, >=stealth, thick},
    ]
    \newcommand{\cellsz}{0.5}
    \newcommand{\depthx}{0.3}
    \newcommand{\depthy}{0.25}
    \coordinate (axisorigin) at (-2.5, -0.25);
    \draw[dimlbl] (axisorigin) -- ++(0.75, 0)
        node[right, font=\small] {$m$ \scriptsize(data)};
    \draw[dimlbl] (axisorigin) -- ++(0, -0.75)
        node[below, font=\small] {$n$ \scriptsize(data)};
    \draw[dimlbl] (axisorigin) -- ++(\depthx, \depthy)
        node[right, font=\small, inner sep=1pt] {$x$ \scriptsize(parallel)};
    \begin{scope}[shift={(0, -0.4)}]
    \begin{scope}[shift={(\depthx, \depthy)}, opacity=0.35]
        \foreach \r/\rr in {0/0,1/1} {
            \foreach \c/\cc in {0/2,1/3} {
                \node[cell] at ({\c*\cellsz}, {-\r*\cellsz}) {$a_{\rr\cc}$};
            }
        }
    \end{scope}
    \foreach \r/\rr in {0/0,1/1} {
        \foreach \c/\cc in {0/0,1/1} {
            \node[cell, fill=white] at ({\c*\cellsz}, {-\r*\cellsz}) {$a_{\rr\cc}$};
        }
    }
    \end{scope}
    \end{tikzpicture}
    \caption{Tensor representation with parallelization dimensions}
    \label{fig:tensor_repr}
\end{figure}

\paragraph{Tensor representation.} We represent how the final output tensors are computed from the input tensors using expressions. Each intermediate tensor computed in a kernel is parallelized across SMs. We use \emph{parallelization dimensions} to represent how the tensors are partitioned. Figure~\ref{fig:tensor_repr} shows a tensor with two data dimensions and one parallelization dimension $x$.

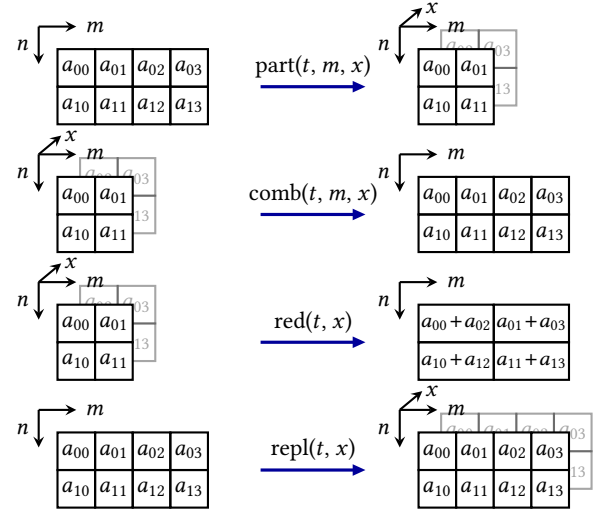
\begin{figure}
    \centering
    \begin{tikzpicture}[
        cell/.style={minimum size=0.5cm, draw, thick, inner sep=0pt, font=\small},
        sumcell/.style={minimum width=0.5cm, minimum height=0.5cm, draw, thick, inner sep=0pt, font=\small},
        arrowlbl/.style={font=\small},
        dimlbl/.style={font=\small, ->, >=stealth, thick},
    ]

    \newcommand{\cellsz}{0.5}
    \newcommand{\depthx}{0.3}
    \newcommand{\depthy}{0.25}
    \newcommand{\leftshift}{-0.5}
    \newcommand{\rightshift}{4.8}
    \newcommand{\rowsep}{-1.7}
    \newcommand{\arrowmid}{3.15}
    \newcommand{\arrowhalf}{0.7}

    \begin{scope}[shift={(\leftshift, 0)}]
        \foreach \r/\rr in {0/0,1/1} {
            \foreach \c/\cc in {0/0,1/1,2/2,3/3} {
                \node[cell] at ({\c*\cellsz}, {-\r*\cellsz}) {$a_{\rr\cc}$};
            }
        }
        \draw[dimlbl] (-0.5, 0.55) -- ++(0.5, 0) node[right, font=\small] {$m$};
        \draw[dimlbl] (-0.5, 0.55) -- ++(0, -0.5) node[left, pos=0.5] {$n$};

        \draw[-{Triangle[length=5pt, width=4pt]}, very thick, blue!60!black] ({\arrowmid-\arrowhalf}, {-0.25}) -- ({\arrowmid+\arrowhalf}, {-0.25});
        \node[arrowlbl] at (\arrowmid, -0.25) [above] {part($t$, $m$, $x$)};

        \begin{scope}[shift={(\rightshift, 0)}]
            \begin{scope}[shift={(\depthx, \depthy)}, opacity=0.35]
                \foreach \r/\rr in {0/0,1/1} {
                    \foreach \c/\cc in {0/2,1/3} {
                        \node[cell] at ({\c*\cellsz}, {-\r*\cellsz}) {$a_{\rr\cc}$};
                    }
                }
            \end{scope}
            \foreach \r/\rr in {0/0,1/1} {
                \foreach \c/\cc in {0/0,1/1} {
                    \node[cell, fill=white] at ({\c*\cellsz}, {-\r*\cellsz}) {$a_{\rr\cc}$};
                }
            }
            \draw[dimlbl] (-0.5, 0.55) -- ++(0.5, 0) node[right, font=\small] {$m$};
            \draw[dimlbl] (-0.5, 0.55) -- ++(0, -0.5) node[left, pos=0.5] {$n$};
            \draw[dimlbl] (-0.5, 0.55) -- ++(\depthx, \depthy)
                node[right, pos=1, font=\small, inner sep=1pt] {$x$};
        \end{scope}
    \end{scope}

    \begin{scope}[shift={(\leftshift, \rowsep)}]
        \begin{scope}[shift={(\depthx, \depthy)}, opacity=0.35]
            \foreach \r/\rr in {0/0,1/1} {
                \foreach \c/\cc in {0/2,1/3} {
                    \node[cell] at ({\c*\cellsz}, {-\r*\cellsz}) {$a_{\rr\cc}$};
                }
            }
        \end{scope}
        \foreach \r/\rr in {0/0,1/1} {
            \foreach \c/\cc in {0/0,1/1} {
                \node[cell, fill=white] at ({\c*\cellsz}, {-\r*\cellsz}) {$a_{\rr\cc}$};
            }
        }
        \draw[dimlbl] (-0.5, 0.55) -- ++(0.5, 0) node[right, font=\small] {$m$};
        \draw[dimlbl] (-0.5, 0.55) -- ++(0, -0.5) node[left, pos=0.5] {$n$};
        \draw[dimlbl] (-0.5, 0.55) -- ++(\depthx, \depthy)
            node[right, pos=1, font=\small, inner sep=1pt] {$x$};

        \draw[-{Triangle[length=5pt, width=4pt]}, very thick, blue!60!black] ({\arrowmid-\arrowhalf}, {-0.25}) -- ({\arrowmid+\arrowhalf}, {-0.25});
        \node[arrowlbl] at (\arrowmid, -0.25) [above] {comb($t$, $m$, $x$)};

        \begin{scope}[shift={(\rightshift, 0)}]
            \foreach \r/\rr in {0/0,1/1} {
                \foreach \c/\cc in {0/0,1/1,2/2,3/3} {
                    \node[cell] at ({\c*\cellsz}, {-\r*\cellsz}) {$a_{\rr\cc}$};
                }
            }
            \draw[dimlbl] (-0.5, 0.55) -- ++(0.5, 0) node[right, font=\small] {$m$};
            \draw[dimlbl] (-0.5, 0.55) -- ++(0, -0.5) node[left, pos=0.5] {$n$};
        \end{scope}
    \end{scope}

    \begin{scope}[shift={(\leftshift, 2*\rowsep)}]
        \begin{scope}[shift={(\depthx, \depthy)}, opacity=0.35]
            \foreach \r/\rr in {0/0,1/1} {
                \foreach \c/\cc in {0/2,1/3} {
                    \node[cell] at ({\c*\cellsz}, {-\r*\cellsz}) {$a_{\rr\cc}$};
                }
            }
        \end{scope}
        \foreach \r/\rr in {0/0,1/1} {
            \foreach \c/\cc in {0/0,1/1} {
                \node[cell, fill=white] at ({\c*\cellsz}, {-\r*\cellsz}) {$a_{\rr\cc}$};
            }
        }
        \draw[dimlbl] (-0.5, 0.55) -- ++(0.5, 0) node[right, font=\small] {$m$};
        \draw[dimlbl] (-0.5, 0.55) -- ++(0, -0.5) node[left, pos=0.5] {$n$};
        \draw[dimlbl] (-0.5, 0.55) -- ++(\depthx, \depthy)
            node[right, pos=1, font=\small, inner sep=1pt] {$x$};

        \draw[-{Triangle[length=5pt, width=4pt]}, very thick, blue!60!black] ({\arrowmid-\arrowhalf}, {-0.25}) -- ({\arrowmid+\arrowhalf}, {-0.25});
        \node[arrowlbl] at (\arrowmid, -0.25) [above] {red($t$, $x$)};

        \begin{scope}[shift={(\rightshift, 0)}]
            \foreach \r/\rr in {0/0,1/1} {
                \node[sumcell, minimum width=1.0cm] at ({0.25}, {-\r*\cellsz})
                    {\footnotesize$a_{\rr 0}\!+\!a_{\rr 2}$};
                \node[sumcell, minimum width=1.0cm] at ({0.25+1.0}, {-\r*\cellsz})
                    {\footnotesize$a_{\rr 1}\!+\!a_{\rr 3}$};
            }
            \draw[dimlbl] (-0.5, 0.55) -- ++(0.5, 0) node[right, font=\small] {$m$};
            \draw[dimlbl] (-0.5, 0.55) -- ++(0, -0.5) node[left, pos=0.5] {$n$};
        \end{scope}
    \end{scope}

    \begin{scope}[shift={(\leftshift, 3*\rowsep)}]
        \foreach \r/\rr in {0/0,1/1} {
            \foreach \c/\cc in {0/0,1/1,2/2,3/3} {
                \node[cell] at ({\c*\cellsz}, {-\r*\cellsz}) {$a_{\rr\cc}$};
            }
        }
        \draw[dimlbl] (-0.5, 0.55) -- ++(0.5, 0) node[right, font=\small] {$m$};
        \draw[dimlbl] (-0.5, 0.55) -- ++(0, -0.5) node[left, pos=0.5] {$n$};

        \draw[-{Triangle[length=5pt, width=4pt]}, very thick, blue!60!black] ({\arrowmid-\arrowhalf}, {-0.25}) -- ({\arrowmid+\arrowhalf}, {-0.25});
        \node[arrowlbl] at (\arrowmid, -0.25) [above] {repl($t$, $x$)};

        \begin{scope}[shift={(\rightshift, 0)}]
            \begin{scope}[shift={(\depthx, \depthy)}, opacity=0.35]
                \foreach \r/\rr in {0/0,1/1} {
                    \foreach \c/\cc in {0/0,1/1,2/2,3/3} {
                        \node[cell] at ({\c*\cellsz}, {-\r*\cellsz}) {$a_{\rr\cc}$};
                    }
                }
            \end{scope}
            \foreach \r/\rr in {0/0,1/1} {
                \foreach \c/\cc in {0/0,1/1,2/2,3/3} {
                    \node[cell, fill=white] at ({\c*\cellsz}, {-\r*\cellsz}) {$a_{\rr\cc}$};
                }
            }
            \draw[dimlbl] (-0.5, 0.55) -- ++(0.5, 0) node[right, font=\small] {$m$};
            \draw[dimlbl] (-0.5, 0.55) -- ++(0, -0.5) node[left, pos=0.5] {$n$};
            \draw[dimlbl] (-0.5, 0.55) -- ++(\depthx, \depthy)
                node[right, pos=1, font=\small, inner sep=1pt] {$x$};
        \end{scope}
    \end{scope}

    \end{tikzpicture}
    \caption{Parallel operators used in \graph verification}
    \label{fig:parallel_ops}
\end{figure}

\paragraph{Parallelization operators.}
Encoding mappings directly at tensor granularity is hard, so we instead decouple them into per-dimension operators. We introduce four parallelization operators (illustrated in Figure~\ref{fig:parallel_ops}):

\begin{compactitem}
    \item $\mathsf{part}(t, m, x)$ (\emph{partition}): Splits data dimension $m$ of tensor $t$ into equal chunks and distributes them across parallel dimension $x$.
    \item $\mathsf{comb}(t, m, x)$ (\emph{combine}): Concatenates the chunks of data dimension $m$ of tensor $t$ across parallel dimension $x$, reconstructing the full dimension. Inverse of $\mathsf{part}$.
    \item $\mathsf{red}(t, x)$ (\emph{reduce}): Performs element-wise sum reduction of tensor $t$ across parallel dimension $x$.
    \item $\mathsf{repl}(t, x)$ (\emph{replicate}): Replicates tensor $t$ across parallel dimension $x$, making identical copies available to each block.
\end{compactitem}

\paragraph{Encoding \graphs.}
We compute the expression of each tensor by traversing the graph in topological order. For each operator, we derive its output expression from its input expressions and the operator semantics. For most operators (e.g., elementwise or matmul), this is straightforward. The key operators are the InputLoader and OutputSaver in the block graph: the InputLoader applies $\mathsf{part}$ or $\mathsf{repl}$ according to the \imap (partitioning the tensor along the mapped dimension, or replicating it if the mapping is $\phi$), and the OutputSaver applies $\mathsf{comb}$ according to the \omap (concatenating the per-block results along the mapped dimension).

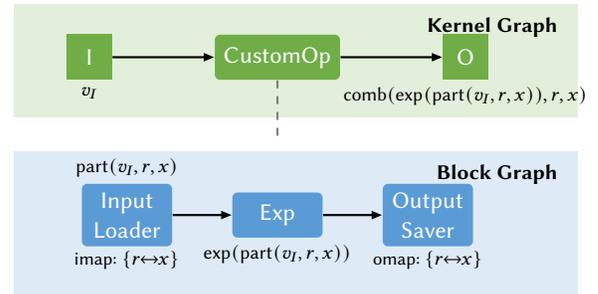
\begin{figure}[h]
    \centering
    \begin{tikzpicture}[
        op/.style={rounded corners=2pt, minimum width=1.2cm, minimum height=0.6cm, font=\small\sffamily, fill=knop_color, text=white},
        subop/.style={rounded corners=2pt, minimum width=1.2cm, minimum height=0.6cm, font=\small\sffamily, fill=tbop_color, text=white},
        tensor/.style={minimum width=0.6cm, minimum height=0.6cm, font=\small\sffamily, fill=knop_color, text=white},
        arr/.style={-{Triangle[length=4pt, width=3pt]}, thick},
        every node/.style={font=\small\sffamily},
    ]
    \fill[kngraph_bg]
        (-1.0, -0.8) rectangle (6.5, 0.7);
    \node[font=\footnotesize\sffamily, anchor=north east] at (6.35, 0.65) {\textbf{Kernel Graph}};

    \node[tensor] (I) at (0, 0) {I};
    \node[font=\scriptsize, anchor=north, yshift=1pt] at (I.south) {$v_I$};
    \node[op] (C) at (2.5, 0) {CustomOp};
    \node[tensor] (O) at (5, 0) {O};
    \draw[arr] (I) -- (C);
    \draw[arr] (C) -- (O);

    \node[font=\scriptsize, anchor=north, yshift=1pt] at (O.south) {$\mathsf{comb}(\mathsf{exp}(\mathsf{part}(v_I, r, x)), r, x)$};

    \draw[dashed, thick, gray] (C.south) -- (2.5, -1.15);

    \fill[tbgraph_bg]
        (-1.0, -1.25) rectangle (6.5, -3.2);
    \node[font=\footnotesize\sffamily, anchor=north east] at (6.35, -1.3) {\textbf{Block Graph}};

    \node[subop, align=center] (In) at (0.5, -2.1) {Input\\Loader};
    \node[font=\scriptsize\sffamily, anchor=north, yshift=2pt] at (In.south) {imap: $\{r\!\leftrightarrow\! x\}$};
    \node[font=\scriptsize, anchor=south, yshift=-1pt] at (In.north) {$\mathsf{part}(v_I, r, x)$};
    \node[subop] (Exp2) at (2.5, -2.1) {Exp};
    \node[font=\scriptsize, anchor=north, yshift=1pt] at (Exp2.south) {$\mathsf{exp}(\mathsf{part}(v_I, r, x))$};
    \node[subop, align=center] (Out2) at (4.5, -2.1) {Output\\Saver};
    \node[font=\scriptsize\sffamily, anchor=north, yshift=2pt] at (Out2.south) {omap: $\{r\!\leftrightarrow\! x\}$};
    \draw[arr] (In) -- (Exp2);
    \draw[arr] (Exp2) -- (Out2);
    \end{tikzpicture}
    \caption{Encoding an \graph as expressions. Each tensor is annotated with its expression; the InputLoader applies $\mathsf{part}$ according to the \imap, and the OutputSaver applies $\mathsf{comb}$ according to the \omap.}
    \label{fig:graph_encoding}
\end{figure}

Figure~\ref{fig:graph_encoding} shows an example. The kernel graph has a single CustomOp that applies elementwise exponentiation with $\imap\!:\{r \leftrightarrow x\}$ and $\omap\!:\{r \leftrightarrow x\}$. The Input operator partitions the input variable $v_I$ along the row dimension over parallel dimension $x$, yielding $\mathsf{part}(v_I, r, x)$. After applying Exp, the OutputSaver combines the result, producing the final expression $\mathsf{comb}(\mathsf{exp}(\mathsf{part}(v_I, r, x)), r, x)$.

\paragraph{Equivalence axioms.}
Table~\ref{tab:rewriting_rules} lists selected equivalence axioms.
The axioms capture algebraic properties of parallelized tensor computation.
We implement equivalence checking using e-graphs~\cite{egg}, which requires converting the axioms to directional rewrite rules.
Some cases require special treatment in the conversion, as discussed further in~\S\ref{sec:impl}.

\begin{table*}[!t]
\centering
\caption{Selected equivalence axioms used in \graph equivalence verification. Notation: $t$ for tensors, $v$ for (batched) vectors, $d$ for data dimensions, $p$ for parallelization dimensions.}
\label{tab:rewriting_rules}
\renewcommand{\arraystretch}{1.25}
\resizebox{\textwidth}{!}{%
\begin{tabular}{@{}ll@{}}
\toprule
\textbf{Axiom} & \textbf{Description} \\

\midrule
\multicolumn{2}{@{}l}{\textit{Matrix multiplication}} \\[2pt]

$\forall\, t_0, t_1, t_2:\quad
  \mathsf{matmul}(t_0,\mathsf{matmul}(t_1, t_2))
  = \mathsf{matmul}(\mathsf{matmul}(t_0, t_1),t_2)$
  & \textsf{matmul} is associative \\

$\forall\, t_0, t_1, t_2:\quad
  \mathsf{matmul}(\mathsf{add}(t_0, t_1),t_2)
  = \mathsf{add}(\mathsf{matmul}(t_0, t_2),\mathsf{matmul}(t_1, t_2))$
  & \textsf{matmul} left-distributes over \textsf{add} \\

$\forall\, t_0, t_1, t_2:\quad
  \mathsf{matmul}(t_0,\mathsf{add}(t_1, t_2))
  = \mathsf{add}(\mathsf{matmul}(t_0, t_1),\mathsf{matmul}(t_0, t_2))$
  & \textsf{matmul} right-distributes over \textsf{add} \\

$\forall\, t_0, t_1, v:\quad
  \mathsf{matmul}(\mathsf{mul}(t_0, v),t_1)
  = \mathsf{mul}(\mathsf{matmul}(t_0, t_1),v)$
  & \textsf{matmul} commutes with \textsf{mul} \\

$\forall\, t_0, t_1, v:\quad
  \mathsf{matmul}(t_0, \mathsf{mul}(t_1, v))
  = \mathsf{mul}(\mathsf{matmul}(t_0, t_1),v)$
  & \textsf{matmul} commutes with \textsf{mul} \\

$\forall\, t_0, t_1, v:\quad
  \mathsf{matmul}(\mathsf{div}(t_0, v),t_1)
  = \mathsf{div}(\mathsf{matmul}(t_0, t_1),v)$
  & \textsf{matmul} commutes with \textsf{div} denominator \\

\midrule
\multicolumn{2}{@{}l}{\textit{Commutativity of parallelization operators}} \\[2pt]

$\forall\, t_0, d_0, p_0, p_1,\; p_0 \neq p_1:\quad
  \mathsf{part}(\mathsf{repl}(t_0, p_0),d_0, p_1)
  = \mathsf{repl}(\mathsf{part}(t_0, d_0, p_1),p_0)$
  & \textsf{part} and \textsf{repl} commute \\

$\forall\, t_0, d_0, p_0, p_1,\; p_0 \neq p_1:\quad
  \mathsf{comb}(\mathsf{repl}(t_0, p_0),d_0, p_1)
  = \mathsf{repl}(\mathsf{comb}(t_0, d_0, p_1),p_0)$
  & \textsf{comb} and \textsf{repl} commute \\

$\forall\, t_0, d_0, p_0, p_1,\; p_0 \neq p_1:\quad
  \mathsf{red}(\mathsf{part}(t_0, d_0, p_1),p_0)
  = \mathsf{part}(\mathsf{red}(t_0, p_0),d_0, p_1)$
  & \textsf{red} and \textsf{part} commute \\

$\forall\, t_0, d_0, p_0, p_1,\; p_0 \neq p_1:\quad
  \mathsf{comb}(\mathsf{red}(t_0, p_0),d_0, p_1)
  = \mathsf{red}(\mathsf{comb}(t_0, d_0, p_1),p_0)$
  & \textsf{comb} and \textsf{red} commute \\

$\forall\, t_0, d_0, d_1, p_0, p_1,\; d_0 \neq d_1,\; p_0 \neq p_1:\quad
  \mathsf{part}(\mathsf{part}(t_0, d_0, p_0),d_1, p_1)
  = \mathsf{part}(\mathsf{part}(t_0, d_1, p_1),d_0, p_0)$
  & Nested \textsf{part} operators commute \\

$\forall\, t_0, d_0, d_1, p_0, p_1,\; d_0 \neq d_1,\; p_0 \neq p_1:\quad
  \mathsf{comb}(\mathsf{comb}(t_0, d_0, p_0),d_1, p_1)
  = \mathsf{comb}(\mathsf{comb}(t_0, d_1, p_1),d_0, p_0)$
  & Nested \textsf{comb} operators commute \\

\midrule
\multicolumn{2}{@{}l}{\textit{Cancellation identities}} \\[2pt]

$\forall\, t_0, d_0, p_0:\quad
  \mathsf{comb}(\mathsf{part}(t_0, d_0, p_0),d_0, p_0)
  = t_0$
  & \textsf{comb} cancels \textsf{part} \\

\midrule
\multicolumn{2}{@{}l}{\textit{Parallelized matrix multiplication}} \\[2pt]

$\forall\, t_0, t_1, p_0:\quad
  \mathsf{red}(\mathsf{matmul}(\mathsf{part}(t_0,\mathit{col}, p_0),
    \mathsf{part}(t_1,\mathit{row}, p_0)),p_0)
  = \mathsf{matmul}(t_0, t_1)$
  & Parallelized \textsf{matmul} (reduction) \\

$\forall\, t_0, t_1, d_0, p_0:\quad
  \mathsf{comb}(\mathsf{matmul}(\mathsf{part}(t_0, d_0, p_0),
    \mathsf{repl}(t_1, p_0)),d_0, p_0)
  = \mathsf{matmul}(t_0, t_1)$
  & Parallelized \textsf{matmul} (row partition) \\

$\forall\, t_0, t_1, d_0, p_0:\quad
  \mathsf{comb}(\mathsf{matmul}(\mathsf{repl}(t_0, p_0),
    \mathsf{part}(t_1, d_0, p_0)),d_0, p_0)
  = \mathsf{matmul}(t_0, t_1)$
  & Parallelized \textsf{matmul} (column partition) \\

$\forall\, t_0, t_1, d_0, p_0,\; d_0 \neq \mathit{row},\; d_0 \neq \mathit{col}:\quad
  \mathsf{comb}(\mathsf{matmul}(\mathsf{part}(t_0, d_0, p_0),
    \mathsf{part}(t_1, d_0, p_0)),d_0, p_0)
  = \mathsf{matmul}(t_0, t_1)$
  & Parallelized \textsf{matmul} (leading dim) \\

\midrule
\multicolumn{2}{@{}l}{\textit{Parallelized sum}} \\[2pt]

$\forall\, t_0, d_0, d_1, p_0,\; d_0 \neq d_1:\quad
  \mathsf{sum}(\mathsf{part}(t_0, d_1, p_0),d_0)
  = \mathsf{part}(\mathsf{sum}(t_0, d_0),d_1, p_0)$
  & \textsf{sum} and \textsf{part} commute \\

$\forall\, t_0, d_0, d_1, p_0,\; d_0 \neq d_1:\quad
  \mathsf{sum}(\mathsf{comb}(t_0, d_1, p_0),d_0)
  = \mathsf{comb}(\mathsf{sum}(t_0, d_0),d_1, p_0)$
  & \textsf{sum} and \textsf{comb} commute \\

$\forall\, t_0, d_0, p_0:\quad
  \mathsf{sum}(\mathsf{repl}(t_0, p_0),d_0)
  = \mathsf{repl}(\mathsf{sum}(t_0, d_0),p_0)$
  & \textsf{sum} and \textsf{repl} commute \\

$\forall\, t_0, d_0, p_0:\quad
  \mathsf{sum}(\mathsf{red}(\mathsf{part}(t_0, d_0, p_0),p_0),d_0)
  = \mathsf{sum}(t_0, d_0)$
  & Compound parallelized \textsf{sum} (form~1) \\

$\forall\, t_0, d_0, p_0:\quad
  \mathsf{sum}(\mathsf{comb}(\mathsf{sum}(\mathsf{part}(t_0, d_0, p_0),d_0),d_0, p_0),d_0)
  = \mathsf{sum}(t_0, d_0)$
  & Compound parallelized \textsf{sum} (form~2) \\

$\forall\, t_0, d_0, p_0, p_1:\quad
  \mathsf{sum}(\mathsf{comb}(\mathsf{red}(\mathsf{part}(\mathsf{part}(t_0, d_0, p_0),d_0, p_1),p_1),d_0, p_0),d_0)
  = \mathsf{sum}(t_0, d_0)$
  & Compound parallelized \textsf{sum} (form~3) \\

$\forall\, t_0, d_0, p_0, p_1:\quad
  \mathsf{sum}(\mathsf{comb}(\mathsf{red}(\mathsf{part}(\mathsf{part}(t_0, d_0, p_0),d_0, p_1),p_0),d_0, p_1),d_0)
  = \mathsf{sum}(t_0, d_0)$
  & Compound parallelized \textsf{sum} (form~4) \\

$\forall\, t_0, d_0, p_0, p_1:\quad
  \mathsf{sum}(\mathsf{red}(\mathsf{comb}(\mathsf{part}(\mathsf{part}(t_0, d_0, p_0),d_0, p_1),d_0, p_0),p_1),d_0)
  = \mathsf{sum}(t_0, d_0)$
  & Compound parallelized \textsf{sum} (form~5) \\

$\forall\, t_0, d_0, p_0, p_1:\quad
  \mathsf{red}(\mathsf{part}(\mathsf{red}(\mathsf{part}(t_0, d_0, p_0),p_0),d_0, p_1),p_1)
  = \mathsf{red}(\mathsf{part}(t_0, d_0, p_0),p_0)$
  & Compound parallelized \textsf{sum} (form~6) \\

$\forall\, t_0, d_0, p_0, p_1:\quad
  \mathsf{red}(\mathsf{red}(\mathsf{part}(\mathsf{part}(t_0, d_0, p_0),d_0, p_1),p_0),p_1)
  = \mathsf{red}(\mathsf{part}(t_0, d_0, p_0),p_0)$
  & Compound parallelized \textsf{sum} (form~7) \\

$\forall\, t_0, d_0, p_0, p_1:\quad
  \mathsf{red}(\mathsf{red}(\mathsf{part}(\mathsf{part}(t_0, d_0, p_0),d_0, p_1),p_1),p_0)
  = \mathsf{red}(\mathsf{part}(t_0, d_0, p_0),p_0)$
  & Compound parallelized \textsf{sum} (form~8) \\

\midrule
\multicolumn{2}{@{}l}{\textit{Parallelized elementwise unary operators}
  $\mathsf{op}_{\mathrm{unary}}$} \\[2pt]

$\forall\, t_0, d_0, p_0:\quad
  \mathsf{part}(\mathsf{op_{\mathrm{unary}}}(t_0),d_0, p_0)
  = \mathsf{op_{\mathrm{unary}}}(\mathsf{part}(t_0, d_0, p_0))$
  & Unary \textsf{op} commutes with \textsf{part} \\

$\forall\, t_0, d_0, p_0:\quad
  \mathsf{comb}(\mathsf{op}_{\mathrm{unary}}(t_0),d_0, p_0)
  = \mathsf{op_{\mathrm{unary}}}(\mathsf{comb}(t_0, d_0, p_0))$
  & Unary \textsf{op} commutes with \textsf{comb} \\

$\forall\, t_0, p_0:\quad
  \mathsf{repl}(\mathsf{op_{\mathrm{unary}}}(t_0),p_0)
  = \mathsf{op_{\mathrm{unary}}}(\mathsf{repl}(t_0, p_0))$
  & Unary \textsf{op} commutes with \textsf{repl} \\

\midrule
\multicolumn{2}{@{}l}{\textit{Parallelized elementwise binary operators }
  $\mathsf{op}_{\mathrm{binary}}$} \\[2pt]

$\forall\, t_0, t_1, d_0, p_0:\quad
  \mathsf{part}(\mathsf{op}_{\mathrm{binary}}(t_0, t_1),d_0, p_0)
  = \mathsf{op}_{\mathrm{binary}}(\mathsf{part}(t_0, d_0, p_0), \mathsf{part}(t_1, d_0, p_0))$
  & Binary \textsf{op} commutes with \textsf{part} \\

$\forall\, t_0, t_1, d_0, p_0:\quad
  \mathsf{comb}(\mathsf{op}_{\mathrm{binary}}(t_0, t_1),d_0, p_0)
  = \mathsf{op}_{\mathrm{binary}}(\mathsf{comb}(t_0, d_0, p_0),\mathsf{comb}(t_1, d_0, p_0))$
  & Binary \textsf{op} commutes with \textsf{comb} \\

$\forall\, t_0, t_1, p_0:\quad
  \mathsf{repl}(\mathsf{op}_{\mathrm{binary}}(t_0, t_1),p_0)
  = \mathsf{op}_{\mathrm{binary}}(\mathsf{repl}(t_0, p_0),\mathsf{repl}(t_1, p_0))$
  & Binary \textsf{op} commutes with \textsf{repl} \\

\bottomrule
\end{tabular}%
}
\end{table*}

Our axioms are intended to be sound---that is, any \graph equivalence derived from the axioms should hold. We note that soundness depends not only on the axioms in isolation but on the entire pipeline, including structural constraints imposed during graph construction. A formal soundness proof is beyond the scope of this paper.
Instead, we rely on careful manual review of the $\sim 70$ axioms,
and also subject all generated kernels to random equivalence testing (which all of them pass).

We do not aim for completeness of the axioms, and known gaps exist, but we do aim to cover all important optimizations. In some cases, this requires instantiating axiom schemata based on some graph features. For example, parallelizing summation across $k$ parallelization dimensions requires depth-$k$ axioms (e.g., the compound parallelized sum axioms in Table~\ref{tab:rewriting_rules}); we enumerate such axioms up to the number of parallelization dimensions in the \graph.
However, there are still cases that our axioms do not aim to cover; for example, computing $T+T$ is the same as multiplying $T$ by the scalar $2$, but this is not covered by our axioms.
The question of whether there exists a recursively enumerable set of axioms that is complete for \graph equivalence is beyond the scope of this paper.

%% file: instantiation.tex
\section{\graph Instantiation}
\label{sec:instantiation}

After verification (\S\ref{sec:optimizer}), we obtain a set of verified \graphs, each with concrete mappings but symbolic parallelization parameters $\mathbf{d} = (d_x, d_i, \ldots)$. The final step is to instantiate these parameters with concrete values that maximize kernel performance for a given input configuration. This is a standard \emph{autotuning} problem: given a set of parameterized kernel templates and a target hardware platform, find the template and parameter values that minimize execution time.

Autotuning for tensor programs and GPU kernels has been extensively studied, with prior work employing a range of strategies including learned cost models~\cite{tvm_auto_tuner, autohalide}, evolutionary search~\cite{ansor}, simulated annealing~\cite{tvm_auto_tuner}, and ensemble methods~\cite{opentuner}. In our setting, the search space consists of the valid values for each parallelization parameter---grid dimension sizes and for-loop iteration counts---subject to the constraint that the resulting tensor dimensions fit within GPU shared memory.

Since kernel compilation dominates the tuning cost, we want to maximize compilation parallelism. To avoid the long dependency chains of iterative methods (e.g., evolutionary search), we adopt random sampling: we uniformly sample valid parameter assignments across all templates, compile and profile them in parallel, and return the best-performing configuration. Integrating more sophisticated tuning strategies is left as future work.

%% file: impl.tex
\section{Implementation}
\label{sec:impl}

We implement \sys on top of the Mirage~\cite{mirage} codebase. The symbolic search components---including the \graph generator, symbolic dimension matching, expression-guided pruning, and mapping instantiation---are implemented in approximately 6,000 lines of C++ (for the symbolic graph representation, search, and verification modules). The e-graph-based equivalence checking for both dimension matching and verification is implemented in approximately 550 lines of Rust using the \texttt{egg} library (version 0.10.0)~\cite{egg}. For kernel code generation, we reuse Mirage's transpiler, which lowers verified $\mu$Graphs into executable CUDA kernels. The symbolic search runs on two Intel Xeon Platinum 8275CL CPUs (48 cores, 96 threads), and discovered kernels are profiled on NVIDIA A100 GPUs.

\paragraph{Adapting axioms for e-graph rewriting.}
E-graph rewriting requires that each rewrite rule $l \to r$ only introduces variables on the right-hand side that already appear on the left-hand side. This means that bidirectional axioms from Table~\ref{tab:rewriting_rules} can only be applied as rewrite rules in the direction that satisfies this constraint.

This constraint poses a challenge when parallelizing multiple operators. For example, the following equivalence for consecutive parallelized matrix multiplications:
\begin{align*}
\mathsf{comb}(&\mathsf{comb}(\mathsf{matmul}(\mathsf{matmul}(\mathsf{repl}(\mathsf{part}(A, r, x), y),\\
&\mathsf{repl}(\mathsf{repl}(B, x), y)), \mathsf{part}(C, c, y)),
 r, x), c, y)\\
=& \mathsf{matmul}(\mathsf{matmul}(A, B), C)
\end{align*}
cannot be verified using only the parallelized matmul axioms (applied left-to-right):
\begin{align*}
&\mathsf{comb}(\mathsf{matmul}(\mathsf{part}(t_0, r, x), \mathsf{repl}(t_1, x)), r, x) \to \mathsf{matmul}(t_0, t_1)\\[2pt]
&\mathsf{comb}(\mathsf{matmul}(\mathsf{repl}(t_0, x), \mathsf{part}(t_1, c, x)), c, x) \to \mathsf{matmul}(t_0, t_1)
\end{align*}
because the nested operators cannot be peeled off one at a time.

To address this, for matmul-related axioms of the form $\mathsf{op}_2(\mathsf{matmul}(\mathsf{op}_0(t_0), \mathsf{op}_1(t_1))) = \mathsf{matmul}(t_0, t_1)$ (where $\mathsf{op}_0, \mathsf{op}_1, \mathsf{op}_2$ are parallelization operators), we additionally introduce the ``inverse'' rewrite rule:
\[
\mathsf{matmul}(\mathsf{op}_0(t_0), \mathsf{op}_1(t_1)) \to \mathsf{op}_2^{-1}(\mathsf{matmul}(t_0, t_1))
\]
where $\mathsf{op}_2^{-1}$ denotes the inverse parallelization operator (e.g., $\mathsf{part}$ for $\mathsf{comb}$ and vice versa). This rule satisfies the variable subset constraint and enables the e-graph to establish equivalences by pushing parallelization operators outward. We apply the same technique to other computation operators that interact with parallelization operators in similar ways.

%% file: eval2.tex
\section{Evaluation}
\label{sec:eval}

\subsection{Experimental Setup}
\label{subsec:eval_setup}

We evaluate \sys by comparing against Mirage's concrete superoptimizer and three additional baselines. \textbf{PyTorch Eager} is standard PyTorch 2.5.1 execution without compilation. \textbf{PyTorch Compiled} uses \texttt{torch.compile} with \texttt{max-autotune} mode, which generates and auto-tunes Triton 3.1.0 kernels. \textbf{TVM (Ansor)}~\cite{ansor} uses Apache TVM 0.18.0 with the Ansor auto-scheduler (1000 tuning trials per workload). For each benchmark, we measure (1) the kernel execution time of the best discovered kernel (compared against all baselines), and (2) the total optimization time (compared against Mirage and TVM, which both involve a search or auto-tuning process). All benchmarks use half-precision floating point. Each kernel is profiled 1,000 times and we report the average execution time.

We evaluate on five workloads commonly found in modern LLMs:
\begin{compactitem}
    \item \textbf{RMSNorm}: $O = \mathsf{matmul}(\mathsf{rms\_norm}(X), W)$, fusing normalization with a linear layer.
    \item \textbf{RMSNorm-MLP}: $O = \mathsf{rms\_norm}(X) \times W_{\mathit{up}} \cdot \mathsf{rms\_norm}(X) \times W_{\mathit{gate}}$, a GLU-style gated MLP with fused normalization.
    \item \textbf{SwiGLU}: $O = \mathsf{silu}(X \times W_{\mathit{gate}}) \cdot (X \times W_{\mathit{up}})$, a gated activation used in LLaMA-style models.
    \item \textbf{Attention}: $O = \mathsf{softmax}(Q \times K^T) \times V$, group-query attention (GQA) in the decode setting.
    \item \textbf{QK-Attention}: $O = \mathsf{softmax}(\mathsf{rms\_norm}(Q) \times K^T) \times V$, GQA with query-key normalization.
\end{compactitem}
For each workload, we evaluate two input configurations with different tensor sizes. For RMSNorm, we vary the hidden dimension $d$ and batch size $n$. For RMSNorm-MLP, we fix $d\!=\!1024$ and vary $n$. For SwiGLU, we fix $n\!=\!8$ and vary $d$. For attention workloads, we fix batch size $b\!=\!2$, number of heads $g\!=\!8$, query sequence length of $1$, and head dimension $d\!=\!128$, and vary the key-value sequence length $h$.

For Mirage, the optimization time consists of (1) graph generation, which enumerates graph structures with concrete mappings, and (2) profiling, which compiles and benchmarks each discovered kernel on the GPU. For \sys, the optimization time consists of (1) symbolic graph generation, which searches graph structures with symbolic mappings, and (2) instantiation, which enumerates valid concrete mapping assignments, verifies each via e-graph equivalence checking, and profiles the best candidates. Note that \sys's graph generation runs once for all input configurations of a workload, while Mirage must search separately for each configuration.

\begin{figure*}
    \centering
    \ifarxiv
    \includegraphics[width=\textwidth]{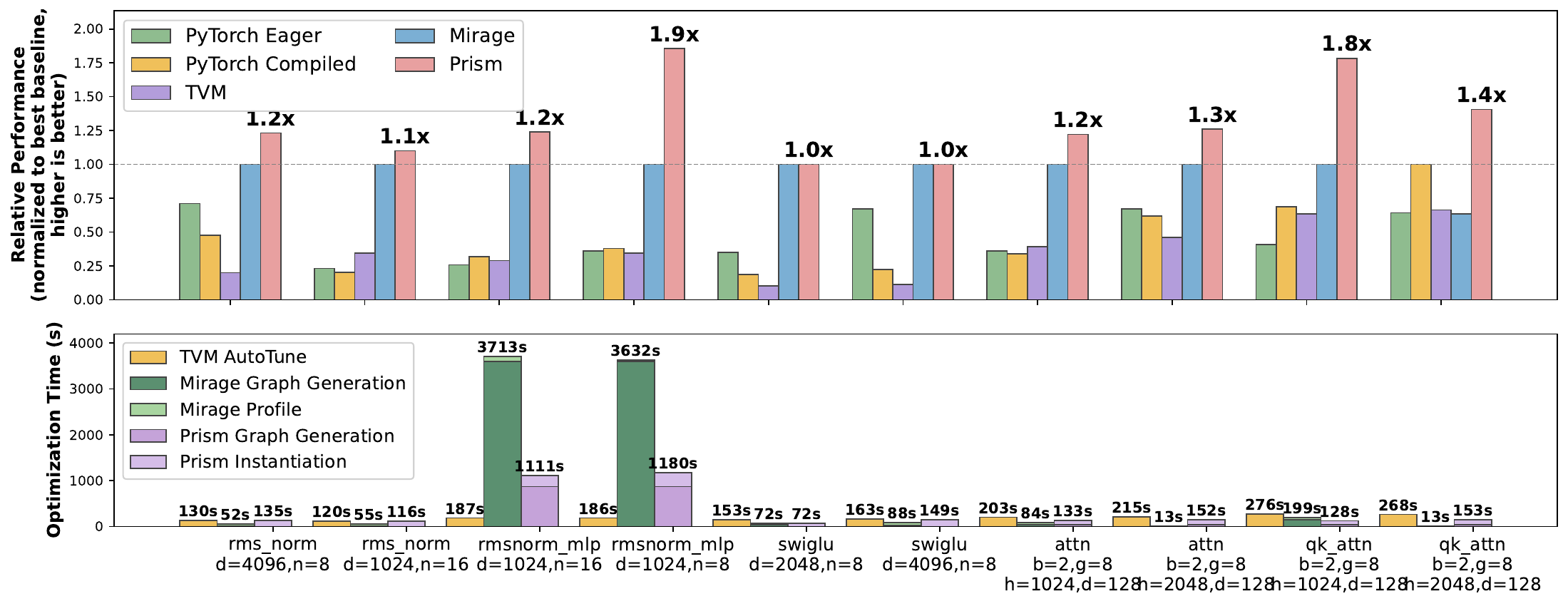}
    \else
    \includegraphics[width=\textwidth]{figs/microbenchmark_eval.pdf}
    \fi
    \caption{Kernel performance and optimization time across 5 workloads. \textbf{Upper}: relative kernel execution time. \textbf{Lower}: total optimization time breakdown---Mirage's time includes graph generation and profiling; \sys's time includes symbolic graph generation and instantiation; TVM's time is the Ansor auto-tuning time (1000 trials).}
    \label{fig:microbenchmark_eval}
\end{figure*}

\subsection{Kernel Performance}
\label{subsec:eval_perf}

Figure~\ref{fig:microbenchmark_eval} compares the kernel execution time and total optimization time of \sys against all baselines. \sys achieves the best kernel time on all 10 configurations, outperforming PyTorch Eager, PyTorch Compiled (torch.compile), TVM (Ansor), and Mirage. For example, on RMSNorm-MLP ($d\!=\!1024, n\!=\!8$), \sys achieves $4.9\times$ speedup over PyTorch Compiled and $5.4\times$ over TVM. The advantage of superoptimization-based approaches over these baselines comes from discovering novel fused kernels that combine multiple operators into a single GPU kernel, reducing memory traffic and kernel launch overhead.

Compared to the existing superoptimizer (Mirage), \sys finds strictly better kernels on 8 configurations and matches on 2 (SwiGLU). The largest improvements are on the attention workloads. Attention involves 3D tensors (batch, sequence, head) that admit many possible parallelization strategies---Mirage uses heuristics to explore only a subset of these mappings and parallelization parameters, while \sys explores the entire mapping and parallelization parameter space symbolically. On QK-Attention, \sys achieves $1.8\times$ ($h\!=\!1024$) and $2.2\times$ ($h\!=\!2048$) speedups over Mirage. On standard attention, the speedups are $1.2\times$ ($h\!=\!1024$) and $1.3\times$ ($h\!=\!2048$). QK-Attention benefits more because the additional normalization operator further expands the space of useful parallelization strategies.

On RMSNorm-MLP, \sys achieves $1.2\times$ ($n\!=\!16$) and $1.9\times$ ($n\!=\!8$) speedups over Mirage---notably, both are configurations where Mirage's concrete search timed out after one hour. On RMSNorm, \sys achieves $1.2\times$ ($d\!=\!4096$) and $1.1\times$ ($d\!=\!1024$) speedups. On SwiGLU, both \sys and Mirage find identical kernels, as the simpler graph structure has fewer mapping choices.

\subsection{Total Optimization Time}
\label{subsec:eval_total}

The lower panel of Figure~\ref{fig:microbenchmark_eval} shows the total optimization time. \sys's total time includes the symbolic search (shared across configurations)---which covers graph generation, mapping enumeration, and verification---plus per-configuration instantiation (parallelization parameter tuning and profiling). Mirage's total time includes per-configuration search plus profiling. TVM's time is the Ansor auto-tuning time with 1000 trials per configuration.

\sys achieves the largest reduction over Mirage on RMSNorm-MLP: Mirage's search times out at one hour (3,713s and 3,632s total), while \sys completes in 1,111s and 1,180s---$3.1\times$--$3.4\times$ faster, while also discovering $1.2\times$--$1.9\times$ faster kernels. On QK-Attention ($h\!=\!1024$), \sys takes 128s vs.\ Mirage's 199s and TVM's 276s, with a $1.8\times$ and $2.8\times$ kernel speedup over Mirage and TVM respectively.

On some configurations, \sys's total time is higher than Mirage's. For RMSNorm ($d\!=\!4096$), \sys takes 135s vs.\ 52s; for attention ($h\!=\!2048$) and QK-Attention ($h\!=\!2048$), \sys takes ${\sim}$152s vs.\ 13s. This is because \sys's instantiation phase has a fixed overhead from compiling and profiling all discovered graph templates, which dominates when Mirage's per-configuration search is already fast. However, \sys still finds better kernels in these cases ($1.2\times$ for RMSNorm, $1.3\times$ and $2.2\times$ for attention and QK-Attention), so the additional optimization time translates into faster end-to-end inference.

\subsection{Search Time Breakdown}
\label{subsec:eval_search}

To better understand the optimization time, Table~\ref{tab:search_time} breaks down the search-only time for \sys and Mirage. For \sys, this includes symbolic graph generation, mapping enumeration, and verification (\S\ref{sec:search}--\S\ref{sec:optimizer}). For Mirage, this includes concrete graph generation with mapping enumeration. \sys's search runs once per workload and covers all input configurations, whereas Mirage searches separately for each configuration.

\begin{table}[t]
    \centering
    \small
    \caption{Search-only time comparison (seconds). \sys's search runs once per workload (shared across all configurations), while Mirage searches per configuration. ``$\times$'' indicates the search was still running at the 1-hour timeout.}
    \label{tab:search_time}
    \begin{tabular}{l|cc|c}
    \toprule
    \multirow{2}{*}{\textbf{Workload}} & \multicolumn{2}{c|}{\textbf{Mirage}} & \textbf{\sys} \\
    & Config 1 & Config 2 & (shared) \\
    \midrule
    RMSNorm       & 11.6s  & 13.0s   & \textbf{0.3s} \\
    RMSNorm-MLP   & 3600s$\times$ & 3600s$\times$ & \textbf{871s} \\
    SwiGLU        & 45.6s  & 27.0s   & \textbf{1.0s} \\
    Attention     & 42.4s  & 10.0s   & \textbf{41s} \\
    QK-Attention  & 154.5s & 10.1s   & \textbf{42s} \\
    \bottomrule
    \end{tabular}
\end{table}

For RMSNorm and SwiGLU, the symbolic search is dramatically faster: 0.3s and 1.0s respectively, compared to Mirage's 11--46s per configuration. The speedup comes from decoupling graph structure search from mapping enumeration---the symbolic search avoids the combinatorial blowup of trying every possible \imap, \fmap, and \omap assignment at each step.

On RMSNorm-MLP, Mirage's concrete search hits the one-hour timeout on both configurations without completing, while \sys finishes in 871s. RMSNorm-MLP fuses two matrix multiplications with normalization and gated multiplication, yielding multiple valid operator orderings that each must be explored for every mapping assignment in the concrete search.

For attention workloads, the search times tell a different story. Attention has a constrained graph structure but a large mapping space due to its 3D tensor structure (batch, sequence, head). Mirage uses heuristics to explore only a subset of the possible mappings and parallelization parameters, resulting in fast per-configuration search (10--42s for attention, 10--155s for QK-Attention). \sys's symbolic search takes 41--42s once for all configurations and explores the full space, which explains why it discovers better kernels (\S\ref{subsec:eval_perf}) despite comparable search times.

\subsection{Graph Diversity}
\label{subsec:eval_diversity}

By exploring the entire mapping and parallelization parameter space symbolically, \sys discovers more unique graphs than Mirage. We consider two graphs unique if they differ in operator sequence or mapping assignments; graphs that differ only in parallelization parameter values are considered the same. Table~\ref{tab:graph_diversity} summarizes the results.

\begin{table}[t]
    \centering
    \small
    \caption{Number of unique graphs discovered by \sys and Mirage (higher is better). Two graphs are unique if they differ in operator sequence or mappings.}
    \label{tab:graph_diversity}
    \begin{tabular}{l|cc|c}
    \toprule
    \multirow{2}{*}{\textbf{Workload}} & \multicolumn{2}{c|}{\textbf{Mirage}} & \textbf{\sys} \\
    & Config 1 & Config 2 & (shared) \\
    \midrule
    RMSNorm       & 1   & 8   & 9 \\
    RMSNorm-MLP   & 12  & 14  & 23 \\
    SwiGLU        & 1   & 1   & 12 \\
    Attention     & 4   & 3  & 14 \\
    QK-Attention  & 4   & 4  & 14 \\
    \bottomrule
    \end{tabular}
\end{table}

\sys discovers 9--23 unique graphs per workload from a single symbolic search, compared to 1--14 per configuration for Mirage. These graphs vary along multiple axes: different numbers of active grid dimensions (1, 2, or 3), different for-loop partitioning strategies, and different operator orderings within the fused kernel.

The difference is most striking on SwiGLU, where Mirage finds only 1 unique structure per configuration while \sys discovers 12, and on attention workloads, where the 3D tensor structure (batch, sequence, head) admits up to 3 grid dimensions. \sys explores all valid combinations of grid and for-loop partitioning across these dimensions, discovering 14 unique graphs for both attention and QK-Attention. Mirage's heuristic-based mapping exploration finds only 3--4 structures per configuration, missing many strategies that \sys discovers. This broader coverage directly translates to the kernel performance improvements observed on attention workloads.

\subsection{Ablation: Impact of Symbolic Maps}
\label{subsec:eval_ablation}

\sys symbolizes three types of mapping variables during search: input maps (\imap), for-loop maps (\fmap), and output maps (\omap). To understand which variables contribute most to the search time reduction, we selectively make individual map types concrete (i.e., enumerated during search) while keeping the rest symbolic. Grid dimension sizes and for-loop range are always deferred to instantiation. Table~\ref{tab:ablation} reports search-only time for RMSNorm ($d\!=\!4096, n\!=\!8$), which has 2 input tensors and 2 data dimensions.

\begin{table}[t]
    \centering
    \caption{Ablation study on RMSNorm: search time when selectively enumerating map variables during search. ``S'' = symbolic (deferred to instantiation), ``C'' = concrete (enumerated during search).}
    \small
    \label{tab:ablation}
    \begin{tabular}{ccc|r}
    \toprule
    \textbf{\imap} & \textbf{\fmap} & \textbf{\omap} & \textbf{Search Time} \\
    \midrule
    S  & S  & S  & \textbf{0.3s} \\
    S  & S  & C & 2.5s \\
    S  & C & S  & 5.5s \\
    C  & S  & S  & 20.5s \\
    S  & C & C & 5.5s \\
    C  & S  & C & 22.5s \\
    C  & C & C & 312s \\
    \bottomrule
    \end{tabular}
\end{table}

Since the mapping search space grows exponentially with the number of parallelization dimensions and data dimensions, symbolizing maps is critical for scalability. When all three map types are enumerated concretely, the search takes 312s; symbolizing all maps reduces this to 0.3s. Among the three map types, \imap contributes the most: enumerating \imap alone yields 20.5s, while enumerating \fmap or \omap alone yields 5.5s and 2.5s respectively. Enumerating multiple map types together further compounds the cost---enumerating all three yields 312s, far exceeding the sum of their individual costs (28.5s).

%% file: related.tex
\section{Related Work}
\label{sec:related}

\paragraph{Expert-crafted kernels.}
A large body of existing systems, including TensorFlow XLA~\cite{tensorflow_xla, Tensorflow}, PyTorch~\cite{pytorch}, and TensorRT~\cite{tensorrt}, depend heavily on kernels engineered by domain specialists for individual ML operators.
In recent years, extensive engineering effort has been invested in refining GPU kernels for widely deployed DNN workloads, especially foundation models~\cite{bommasani2022opportunities}.
Attention mechanisms~\cite{transformers}, for instance, have seen a sequence of highly tuned implementations derived from FlashAttention~\cite{dao2023flash, tri2023flashdecoding, hong2024flashdecoding, fastertransformer}.
However, the rapid evolution of GPU architectures (e.g., the introduction of tensor cores in A100 GPUs~\cite{Markidis2018tensorcores}, thread block clusters in H100 GPUs~\cite{nvidia-h100}, and tensor memory in B200 GPUs) substantially enlarges the optimization space. 
As a result, manually engineered kernels are increasingly prone to overlooking non-obvious performance opportunities that are difficult to identify through human-driven design alone.

\paragraph{Superoptimization-based methods.}
Superoptimization was originally proposed to derive optimal instruction sequences automatically~\cite{massalin, stoke, peephole}, and has since been extended to tensor program optimization~\cite{TASO, wang2021pet, zheng2023einnet, tensat, unger2022unity, FlexFlow, hu2024korch, jeon2025graphpipe}. Systems such as TASO~\cite{TASO} enumerate equivalent subgraphs with correctness validated via testing and formal methods, while Mirage~\cite{mirage} expands the search to multiple levels of the GPU execution hierarchy. More recent approaches, such as AlphaEvolve~\cite{novikov2025alphaevolve}, leverage LLM-guided evolutionary search to explore larger optimization spaces.

These methods fall into two categories: enumeration-based techniques (e.g., TASO, Mirage), which provide structured but poorly-scalable search, and sampling-based techniques (e.g., AlphaEvolve), which scale but lack coverage guarantees. In contrast, \sys introduces a symbolic superoptimization framework that compactly represents families of tensor programs and enables sound pruning of the search space while preserving optimal solutions.

\paragraph{Symbolic graph representations.} Prior work has introduced methods to represent tensor programs using multi-level graph representations. For example, Welder~\cite{shi2023welder} and ASPEN~\cite{park2023aspen} use a tile-based, multi-level graph to represent tensor programs. Mirage~\cite{mirage} introduces a multi-level graph representation to capture the GPU hierarchy. Unlike these approaches that represent concrete tensor programs, \sys leverages a symbolic, hierarchical representation of tensor programs to compactly encode large equivalence classes of tensor programs to reduce the search space.